\documentclass[12pt]{article}
\usepackage{amsmath,amssymb}
\usepackage{a4wide}
\usepackage{epsfig}
\newcommand{\I}{\text{i}}
\newcommand{\E}{\text{e}}

\newcommand{\STr}{\text{STr}}
\newcommand{\Tr}{\text{Tr}}
\newcommand{\trc}{\text{tr}_{\text{c}}}

\newcommand{\re}[1]{~(\ref{#1})}
\newcommand{\case}[2]{{\scriptstyle \frac{#1}{#2}}}
\newcommand{\Gk}{\Gamma_k}
\newcommand{\Gi}{\Gamma_k^{\text{inv}}}
\newcommand{\Ggh}{\Gamma_k^{\text{gh}}}
\newcommand{\Ggf}{\Gamma_k^{\text{gf}}}
\newcommand{\Ga}{\Gamma_k^{\text{gauge}}}
\newcommand{\Gt}{\Gamma_k^{(2)}}
\newcommand{\Aa}{\mathsf{A}}
\newcommand{\Fa}{\mathsf{F}} 
\newcommand{\bA}{\bar{A}}
\newcommand{\tet}{\theta}
\newcommand{\te}{\vartheta}
\newcommand{\wdot}{\dot{w}_k}
\newcommand{\wddot}{\ddot{w}_k}
\newcommand{\gb}{\bar{g}}
\newcommand{\Bl}{\bar{B}_l}
\newcommand{\ZFk}{Z_{\text{F},k}}

\newcommand{\hi}{\tilde{h}(s)}
\newcommand{\gt}{\tilde{g}(s)}
\newcommand{\Hl}{\widetilde{H}(s)}
\newcommand{\Gl}{\widetilde{G}(s)}
\newcommand{\Dl}{{\cal D}_{\text{L}}}
\newcommand{\Dt}{{\cal D}_{\text{T}}}

\newcommand{\pd}{\partial}
\newcommand{\pat}{\partial_t}
\newcommand{\Pl}{P_{\text{L}}}
\newcommand{\Pt}{P_{\text{T}}}
\newcommand{\Pp}{P_{\|}}
\newcommand{\Pb}{P_{\bot}}
\begin{document}

$\text{}$

\vspace{-2cm}

{\hfill CERN-TH/2002-047}

\vspace{1.5cm}

\centerline{\Large\bf Running coupling in Yang-Mills theory}
\centerline{\Large\bf  -- a flow equation study -- } 

\vspace{.8cm}

\centerline{\large Holger Gies}
 
\vspace{.6cm}

\centerline{\small\it CERN, Theory Division, CH-1211 Geneva 23,
  Switzerland }
\centerline{\small\it \quad E-mail: Holger.Gies@cern.ch}

\begin{abstract}
  The effective average action of Yang-Mills theory is analyzed in the
  framework of exact renormalization group flow equations. Employing
  the background-field method and using a cutoff that is adjusted to
  the spectral flow, the running of the gauge coupling is obtained on
  all scales. In four dimensions and for the gauge groups SU(2) and
  SU(3), the coupling approaches a fixed point in the infrared. 
\end{abstract}

\section{Introduction and Summary}

Understanding the infrared sector of Yang-Mills theory still
represents a challenge in quantum field theory. The strong coupling of
the system and the rich dynamics of its degrees of freedom are well
beyond the applicability of many field theoretic methods. Even without
attempting to solve the theory at one fell swoop, it is already
difficult to find (and then answer) questions that can be disentangled
from the full complexity of the problem.

In this work, we study Yang-Mills theory in the framework of
renormalization group (RG) flow equations \cite{Wegener:1972ih} for
the effective average action \cite{Wetterich:1993yh}, concentrating
solely on the running gauge coupling. Whereas perturbation theory
describes asymptotic freedom of the coupling in the high-energy limit,
it fails to predict anything at low energies except for its own
failure -- manifested by the Landau pole singularity. Even without
unveiling the complete infrared structure of gauge theories (including
confinement and a mass gap), an analytic knowledge of the running of
the coupling towards lower energies beyond perturbation theory is
desirable. Exact RG flow equations represent an appropriate tool for
tackling this problem. \bigskip

\noindent
{\bf Flow equation for the effective average action.}
Being a ``coarse-grained'' free-energy functional, the effective
average action $\Gk$ governs the dynamics of a theory at a momentum
scale $k$.  It comprises the effects of all quantum fluctuations of
the dynamical field variables with momenta larger than $k$, whereas
fluctuations with momenta smaller than $k$ have not (yet) been
integrated out.  Decreasing $k$ corresponds to integrating out more
and more momentum shells of the quantum fluctuations. This successive
averaging is implemented by a $k$-dependent infrared cutoff term
$\Delta_k S$ which is added to the classical action in the standard
Euclidean functional integral. This term gives a momentum-dependent
mass square $R_k(p^2)$ to the field modes with momentum $p$ which
vanishes for $p^2\gg k^2$.  Regarding $\Gk$ as a function of $k$, the
effective average action runs along a RG trajectory in the space of
all action functionals that interpolates between the classical action
$S=\Gamma_{k\to \infty}$ and the conventional quantum effective action
$\Gamma=\Gamma_{k\to0}$. The response of $\Gk$ to an infinitesimal
variation of the scale $k$ is described by a functional differential
equation, the flow equation (exakt RG equation). In a symbolic
notation,
\begin{equation}
\pat\Gk=\frac{1}{2}\, \STr\,\Bigl[ \pat R_k\, \bigl(\Gt+R_k\bigr)^{-1}
\Bigr],\quad \pat\equiv k\frac{d}{dk}, \label{I.1}
\end{equation}
where $\Gt$ denotes the second functional derivative of the effective
average action with respect to the field variables and corresponds to
the inverse exact propagator at the scale $k$. \bigskip

\noindent
{\bf Flow equation in gauge theories.}
The use of flow equations in gauge theories, as initiated in
\cite{Bonini:1993sj}, \cite{Ellwanger:iz}, \cite{Reuter:1993kw}, is
complicated by the fact that it is difficult to reconcile the
Wilsonian idea of integrating out momentum shells of quantum
fluctuations with gauge invariance.  Working with gauge-noninvariant
field variables such as gluons and ghosts, a regularization of the
theory with a momentum cutoff necessarily breaks gauge invariance.
Nevertheless, gauge-invariant flows can, in principle, be constructed
by taking care of constraints imposed by the Ward identities which are
modified by the presence of the cutoff \cite{Ellwanger:iz},
\cite{Bonini:1994kp}, \cite{Ellwanger:1995qf},
\cite{D'Attanasio:1996jd}; in practice, resolving these constraints
beyond perturbation theory is highly involved; for a review, see
\cite{Litim:1998nf}.

As an alternative, a formulation in terms of gauge-invariant variables
such as, for instance, Wilson loops may therefore be desirable and so
has been proposed and worked out in \cite{Morris:1999px}. Related to
this, a gauge-invariant regularization has been formulated in
\cite{Morris:2000fs} by constructing SU($N$) Yang-Mills theory from a
spontaneously broken SU($N|N$) super-gauge extension; here the
fermionic super-partners become massive and act as Pauli-Villars
regulator fields without breaking the residual SU($N$) gauge
invariance. As a result, the one-loop $\beta$ function has been
computed without any gauge fixing. 

In this work, we decide to employ the conventional and technically
more feasible formulation in terms of the gluonic gauge field at the
expense of only partially resolving the modified Ward identities
resulting in less control over gauge invariance. In this way, we
shall accept a compromise between calculational advantages and the
implementation of complete quantum gauge invariance. In particular, we
follow the strategy of \cite{Reuter:1997gx}, employing the
background-field method. Our solution to the flow equation will be
gauge invariant in the background field, but the renormalization group
trajectory that connects the classical (bare) action with our quantum
solution will not satisfy all requirements of gauge invariance
(cf.~Sect.~\ref{setting}). \bigskip

\noindent
{\bf Truncations.}
Flow equations for interacting quantum field theories can be solved
only approximately. A consistent and systematic approximation scheme
is given by the method of truncations. Herein, the infinite space of
all possible actions, spanned by the field operators compatible with
the symmetries, is truncated to a subset of operators; the flow
equation for the complete effective action can then be boiled down to
the flow equations of the coefficients of these operators (generalized
couplings). The renormalization trajectory in the space of all actions
is thereby projected onto the hypersurface spanned by all operators of
the truncation. For a selected truncation to be able to describe the
physics of the system, its operators have to cover the dynamics of the
relevant degrees of freedom of the system under consideration. Since
the relevant degrees of freedom in strongly coupled quantum field
theories such as Yang-Mills theories may change under the
renormalization flow, a careful and deliberate choice of the
truncation is halfway to the solution of the theory. In view of the
many proposals concerning the ``true'' degrees of freedom in the
infrared sector of Yang-Mills theory, their systematic study within a
flow equation approach would be desirable. Along this direction,
interesting and promising results have been obtained in
\cite{Ellwanger:1999vc} and \cite{Freire:2001nd}, where the choices of
the truncation have been based on the monopole picture of infrared
Yang-Mills theory.

In the present work, we follow a different strategy: we stick to the
``gluonic language'' and maintain the gauge field as the basic
variable. This avoids complications inherent in the change of quantum
variables, which has to be performed with great care (see, e.g.,
\cite{Latorre:2000qc} and \cite{Gies:2001nw}). But in order to account
for the fact that the ``true'' infrared degrees of freedom may have a
complicated gluonic description, we include infinitely many gluonic
invariants in our truncation; to be explicit, we consider a truncation
in which the gauge-invariant part of the effective action is an
arbitrary function $W_k$ of the square of the field strength $F$,
\begin{equation}
\Gi[A]=\int W_k(\tet), \quad \tet:=\frac{1}{4} {
  F}_{\mu\nu}^a F_{\mu\nu}^a, \label{I.2}
\end{equation}
and the running of the coupling will be extracted from the flow of the
linear $F_{\mu\nu}^a F_{\mu\nu}^a$ term in $W_k$, as it is standard in
continuum quantum Yang-Mills theory. At weak coupling, it may be
sufficient to approximate $W_k[\tet]$ by a finite series, i.e., a
polynomial in $\tet$, being justifiable by simple power counting
(higher operators are suppressed by powers of the ultraviolet cutoff).
But at strong coupling, those higher operators can acquire large
anomalous dimensions that completely obstruct a naive power-counting
analysis.  In fact, our results show that the flow of the complete
function $W_k$ contributes to the running gauge coupling, and that the
flow of higher order operators must not be neglected.

Beyond the approximations involved (i) in choosing Eq.\re{I.2} as our
truncation (and neglecting other invariants) and (ii) in resolving the
modified Ward identity only partially, we make a third approximation
(iii) by neglecting any nontrivial running in the ghost and gauge-fixing
sectors. \bigskip

\noindent
{\bf Regulators.}
For an explicit evaluation of the flow equation, a cutoff function (or
regulator) $R_k$ has to be specified. This cutoff function is to some
extent arbitrary (see Sect.~\ref{cutoff}). In the denominator of the
flow equation\re{I.1}, it acts as an infrared cutoff for modes with
momenta smaller than $k$; its derivative $\pat R_k$ in the numerator
is peaked $\delta$-like around $k$ and thus implements the Wilsonian
idea of integrating successively over momentum shells. Different
choices of $R_k$ correspond to different RG trajectories in the space
of all action functionals. But by construction, the complete quantum
solution $\Gamma=\Gamma_{k\to 0}$, being the endpoint of all
trajectories, is independent of $R_k$.

This $R_k$ independence of the solution, of course, holds only for
exact solutions to the flow equation. Approximations such as the
choice of a truncation generically introduce a cutoff dependence of
the final result. On the one hand, this is clearly a disadvantage of
the method; one is led to study one and the same problem with many
different cutoffs in order to extract cutoff-independent information.
On the other hand, after having accepted that exact solutions might
never be at our disposal for most quantum field theories, we can
exploit the cutoff dependence in order to improve our approximations.

In order to illustrate this point, let us recall that truncations cut
a hypersurface out of the space of all action functionals. A
truncation will be acceptable if the complete quantum effective
action lies within or close to this hypersurface. But this is not
a sufficient criterion: imagine a certain exact RG trajectory
(corresponding to a certain cutoff function) that begins and ends
within this hypersurface, but in between develops a large distance to
the hypersurface%, as sketched in Fig.\ref{fig1}
. In the exact theory,
this flow may largely be driven by operators which do not belong to the
truncation spanning the hypersurface. Working only within the
truncation, the contribution of these other operators cannot be
accounted for, and the so-found solution to the flow will generally be
different from the true solution.

Instead, the optimal strategy would be to choose those exact RG
trajectories (and their corresponding cutoff functions) that lie
completely in (or close to) the hypersurface. But strictly speaking,
this ideal case is not possible, since the cutoff function generally
couples the flow to all operators, so that an RG trajectory will never
lie only within a restricted hypersurface. A more precise criterion
would be that the truncated RG trajectory within the hypersurface
should be equal to (or close to) the exact RG trajectory after
projecting the latter onto the hypersurface. Then, the flow towards
the quantum solution is driven mainly by the operators contained in
the truncation, and the final result will represent a good
approximation to the exact one. However, we are currently not aware of
any method that fully formalizes these ideas. Up to now, the
properties of the flow that depend on the cutoff function can only be
investigated within a given truncation.

However, a systematic study of cutoff functions has recently been
put forward mainly within derivative-expansion truncations in scalar
and fermionic theories, and ``optimized'' cutoff functions have been
proposed \cite{Litim:2000ci}. The optimization criterion focuses on
improving the convergence of approximate solutions to flow equations;
in fact, for scalar O($N$) symmetric theories, it leads to better
results for the critical exponents \cite{Litim:2001dt}. \bigskip

\noindent
{\bf Spectrally adjusted cutoff.}
The class of cutoff functions employed in this work is also considered
to be improved in the sense mentioned above. In this case, the
improvement does not refer to the precise shape of the cutoff
function, but rather to the choice of its argument. Here, we will use
not just the spectrum of the Laplace operator (which would be the
gauge-covariant generalization of the momentum squared), but the full
second functional derivative of the effective average action $\Gt$
evaluated at the background field.

The argument of the cutoff function can be understood as a parameter
which controls the order and size of the momentum shell that is
integrated out upon lowering the scale from $k$ to $k-\Delta k$. It
appears natural that a truncated flow can be controlled better if
each momentum shell covers an equal part of the spectrum of quantum
fluctuations. The spectrum itself is not fixed, but $k$ dependent;
lower modes get dressed by integrating out higher modes. In order
to adapt the cutoff function to this spectral flow, we insert the full
$\Gt$ into its argument, and so obtain a ``spectrally adjusted''
cutoff.

This has two technical consequences: first, as the flow equation is
evaluated at the background field in our truncation, the right-hand
side can be transformed into a propertime representation; here, we
have powerful tools at our disposal that allow us to keep track of the
full dependence of the flow equation on the field strength
squared. Secondly, the degree of nonlinearity of the flow equation
strongly increases, inhibiting its straightforward analytical or 
numerical computation even within simple truncations. We solve
this technical problem by first expanding the flow for the gauge
coupling in an asymptotic series, and then reconstructing an integral
representation for this series by analyzing the leading (and
subleading) asymptotic growth of the series coefficients. Whereas
most parts of our work are formulated in $d>2$ dimensions and for the
gauge group SU($N$), this final analysis concentrates on the most
interesting cases of $d=4$ and $N=2$ or $N=3$. \bigskip
 
\begin{figure}
\begin{center}
\begin{picture}(120,57)
\put(-8,-10){
\epsfig{figure=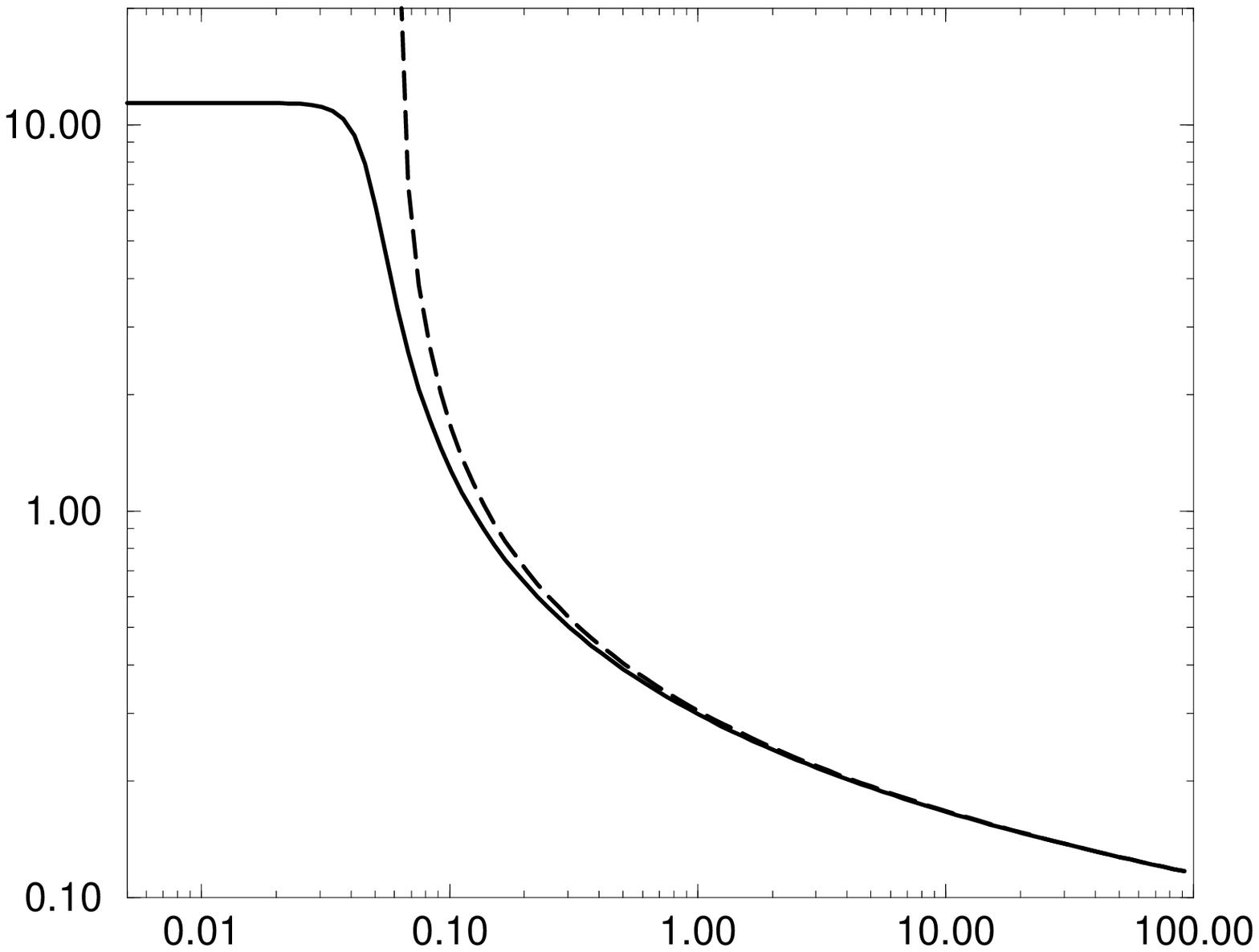,width=13cm,height=8cm}}
\put(5,61){$\alpha_{\text{s}}$}
\put(90,53){SU(2)} 
\put(110,-8){$k/$GeV} 
\put(42,55){1-loop}
\end{picture} 
\end{center} 
\caption{Running coupling $\alpha_{\text{s}}$ versus momentum scale
  $k$ in GeV for gauge group SU(2), using the initial value
  $\alpha_{\text{s}}(M_Z)\simeq0.117$. The solid line represents the
  result of our calculation in comparison with one-loop perturbation
  theory (dashed line).}
\label{figalpha} 
\end{figure}

\noindent
{\bf Results.}
As a result, we find a representation of the $\beta$ function of
Yang-Mills theory. For weak coupling, we rediscover an
accurate perturbative behavior. As the scale $k$ approaches the
infrared, the coupling grows and finally tends to an infrared stable
fixed point, $\alpha_{\text{s}}\to\alpha_\ast$. Our quantitative
results are
\begin{eqnarray}
\alpha_\ast&\simeq& 11.3 \quad \text{for SU(2)}, \nonumber\\
\alpha_\ast&\simeq& 7.7\pm 2 \quad \text{for SU(3)}. \label{I.3}
\end{eqnarray}
The uncertainty in the SU(3) case arises from an unresolved color
structure in our calculation (cf.~App.~\ref{SU3}).

The complete flow of the running coupling is depicted in
Fig.~\ref{figalpha} for pure SU(2) Yang-Mills theory in comparison
with perturbation theory. For illustrative purposes, we use
$\alpha_{\text{s}}(M_Z)\simeq0.117$ as initial value ($M_Z\simeq 91.2$
GeV). Sizeable deviations from perturbation theory occur for
$k\lesssim 1$ GeV, and the fixed point plateau is reached for $k={\cal
  O}(10\text{MeV})$. We shall argue below that a larger truncation as
well as the inclusion of dynamical quarks are expected to decrease the
value of $\alpha_\ast$.

The paper is organized as follows: Sect.~\ref{setting} briefly recalls
the framework of flow equations in gauge theories with the
background-field method and describes our basic approximations. In
Sect.~\ref{floweq}, we boil down the flow equation as required for our
truncation. Sect.~\ref{expansion} is devoted to extracting the RG flow
of the running gauge coupling, which is the main result of the present
work. The role of the spectrally adjusted cutoff is illustrated in
Sect.~\ref{role}. Sect.~\ref{conclusions} contains our conclusions and
a discussion of our results in the light of related literature.

\section{Flow equation for Yang-Mills theory}
\label{setting}

We begin with a brief outline of the flow equation and the
background-field formalism as they are employed in this work. We focus
on direct applicability and the required approximations and leave
aside more formal (though important) aspects, as they are presented in
\cite{Reuter:1997gx} and \cite{Freire:2000bq}. Let us therefore start
with a more explicit representation of the flow equation for the
effective average action, 
\begin{equation}
\pat \Gk[A,\bA] = \frac{1}{2}\, \STr\, \bigg\{\pat R_k(\Gt[\bA,\bA])\,\,
\Big[ \Gt[A,\bA]+R_k(\Gt[\bA,\bA]) \Big]^{-1}\bigg\}, \label{2.5}
\end{equation}
where we denote the so-called classical gauge field by ${ A}_\mu^a$,
which is the usual field variable of the quantum effective action
(conjugate to the source). We also introduce a background field
$\bA_\mu^a$, and have already inserted $\Gt$ evaluated at the
background field into the cutoff function.\footnote{This $\Gt$ is
  evaluated at the background field because an $A$ dependence would
  spoil the one-to-one correspondence of the flow equation to the
  functional integral.} The symbol STr implies tracing over all
internal indices and provides for a minus sign in the ghost sector.
We aim at solving Eq.\re{2.5}, using the following truncation:
\begin{equation}
\Gk[A,\bA]=\Gi[A]+\Ggf[A,\bA]+\Ggh[{
  A},\bA]+\Ga[A,\bA]. \label{2.1} 
\end{equation}
Following \cite{Abbott:1980hw}, the background-field method is
introduced to enable us not only to perform a meaningful integration over
gauge-fixed quantum fluctuations but to simultaneously arrive at a
gauge-invariant effective action. Identifying the quantum fluctuations
with $A-\bA$, the gauge-fixing term
\begin{equation}
\Ggf[A,\bA]=\frac{1}{2\alpha} \int_x
\big[D_\mu[\bA]\,(A-\bA)_\mu\big]^2 \label{2.2}
\end{equation}
with a gauge parameter $\alpha$ is invariant under a simultaneous
gauge transformation of $A_\mu^a$ and $\bA_\mu^a$, and so is the ghost
action
\begin{equation}
\Ggh[A,\bA]=-\int_x \bar{c}\, D_\mu[\bA]D_\mu[A]\, c,\label{2.3}
\end{equation}
where the ghosts $\bar{c},c$ are understood to transform
homogeneously. We should stress that with this truncation of the ghost
and gauge-fixing sector, we neglect any running there. 

If we solved the theory completely, the resulting quantum effective
action $\Gamma_{k=0}[A,\bA]$ would be gauge invariant precisely
at $A=\bA$. Imposing a normalization of $\Ga[A,\bA]$ such
that
\begin{equation}
\Ga[A=\bA,\bA]=0,\label{2.4}
\end{equation}
we conclude that the so-found solution $\Gamma^{\text{inv}}_{k\to
  0}[A]$ would be gauge invariant and would represent the desired
quantum effective action (provided that we also worked out the
complete ghost sector). The quantity $\Ga$ hence parametrizes the
gauge-noninvariant remainder of the action in the physically
  irrelevant case of $A\neq\bA$. 

Although we are finally interested in $\Gk[A=\bA,\bA]$, $\Ga$
cannot be dropped right from the beginning in Eq.\re{2.5}, because its
second functional derivative $(\Ga)^{(2)}[A,\bA]\neq 0$ in general;
$\Ga$ contributes to the flow of $\Gk[A,\bA]$ even at ${
  A}=\bA$. Still, neglecting $\Ga$ seems to be a consistent
truncation if we are interested only in $\Gi[A=\bA]$. But besides
Eq.\re{2.5}, the effective action also has to satisfy the constraints
imposed by gauge invariance in the form of the modified Ward
identity; in symbolic notation,
\begin{equation}
{\cal L}_{\text{W}}[\Gk]=\Delta[R_k],\label{2.6}
\end{equation}
where ${\cal L}_{\text{W}}$ denotes the usual Ward operator
constraining the effective action $\Gk$, and the cutoff-dependent
right-hand side represents the modification due to the infrared
regulator $R_k$ (for an explicit representation of Eq.\re{2.6}, see
\cite{Reuter:1997gx}, \cite{Freire:2000bq}). In fact, if the cutoff is
removed in the limit $k\to 0$, we rediscover the standard Ward
identity ${\cal L}_{\text{W}}[\Gk]=0$. Inserting our truncation\re{2.1}
into the Ward identity, the first three terms drop out and we are left
with
\begin{equation}
{\cal L}_{\text{W}}[\Ga]=\Delta[R_k].\label{2.7}
\end{equation}
This tells us that, on the one hand, $\Gi$ is indeed not constrained
by the modified Ward identity and any gauge-invariant ansatz is
allowed; on the other hand, a vanishing $\Ga$ is generally
inconsistent with the constraint. It is a nontrivial assumption of
this work that $\Ga$ as driven by the right-hand side of Eq.\re{2.7}
does not strongly influence the flow of $\Gi$ at $A=\bA$, so that we
can safely neglect it in a first approximation.

With regard to our final asymptotic analysis of the running coupling,
we can even weaken this assumption a bit: since we reconstruct the
$\beta$ function from its asymptotic series expansion by analyzing its
leading growth, neglecting $\Ga$ corresponds to assuming that $\Ga$
does not strongly modify this leading growth. In view of the fact that
$\Ga$ starts from zero in the ultraviolet and enters the flow only
indirectly, this assumption appears rather natural, at least for a
large part of the flow.

We should remark that the effective average action $\Gk$ has to
satisfy another identity that can be derived by considering the
response of $\Gk$ on gauge transformations of the background field
only. This background-field identity is in close relation to the
modified Ward identity \cite{Freire:2000bq} (for an explicit proof in
QED, see\cite{Freire:1996db}), and also imposes a constraint only on
$\Ga$ similar to Eq.\re{2.7}. As has been shown in
\cite{Freire:2000bq}, this identity does not cause further fine-tuning
problems which would add to those that are posed by the modified Ward
identity.

In summary, solving the flow equation\re{2.5} with the
truncation\re{2.1} will result in an action functional $\Gi[{
  A}=\bA]$ which is invariant under the background-field
transformation. By neglecting $\Ga$, this invariance is not identical
to full quantum gauge invariance even at $A=\bA$, since the flow
is not completely compatible with the modified Ward identities. This
work is based on the assumption that these violations of
quantum gauge invariance have little effect on the final result.

\section{Evaluation of the truncated flow}
\label{floweq}
We shall now solve the flow equation\re{2.5} within the
truncation\re{2.1} (neglecting $\Ga$) and with $\Gi$ as given in
Eq.\re{I.2},
\begin{equation}
\Gi[A]=\int_x\, W_k(\tet), \quad W_k(\tet)=\sum_{i=1}^{\infty}
\frac{W_i}{i!}\, \tet^i, \label{3.1}
\end{equation}
where $\tet:=\frac{1}{4} { F}_{\mu\nu}^a F_{\mu\nu}^a$. An important
ingredient of the flow equation is the cutoff function $R_k$, which we
display as
\begin{equation}
R_k(x)=x\, r(y), \quad y:=\frac{x}{Z_k k^2},\label{3.2}
\end{equation}
with $r(y)$ being a dimensionless function of a dimensionless
argument. We include wave-function renormalization constants $Z_k$ in
the argument of $r(y)$ for reasons to be discussed below. Note that
$Z_k$ as well as $R_k$ itself are matrices in field space; different
field variables may be accompanied by different $Z_k$'s and $R_k$'s.
The cutoff function $R_k$ has to satisfy the following standard
constraints:
\begin{equation}
\lim_{x/k^2\to 0} R_k(x)>0,\quad\lim_{k^2/x\to 0} R_k(x)=0, \quad
\lim_{k\to\Lambda} R_k(x)\to \infty,\label{3.3}
\end{equation}
which guarantee that $R_k$ provides for an infrared regularization,
ensure that the regulator is removed in the limit $k\to 0$, and
control the ultraviolet limit where $\Gk{}_{\to\Lambda}=S_\Lambda$
should approach its initial condition $S_\Lambda$ at the initial
ultraviolet scale $\Lambda$. These constraints are met by the
representation\re{3.2} and translate into constraints for $r(y)$.
Since we will identify the argument $x$ with the full $\Gt$ at the
background field, the first constraint of\re{3.3} must be formulated
more strongly,
\begin{equation}
\lim_{x/k^2\to 0}  R_k(x)=Z_k\,k^2, \quad r(y\to 0) \to \frac{1}{y},
\label{3.4}
\end{equation}
in order to guarantee that the one-loop approximation of the flow
equation results in the true one-loop effective action. We shall not
specify $r(y)$ any further until we employ an exponential cutoff for
the final quantitative computation (see. Eq.\re{D4}).

Within the approximations mentioned above, the flow equation\re{3.5}
can be written as
\begin{eqnarray}
\pat\Gk[A=\bA,\bA]&=&\frac{1}{2}\, \STr\,\frac{\pat
  R_k(\Gt)}{\Gt+R_k(\Gt)} \nonumber\\
&=&\frac{1}{2}\, \STr \left[ (2-\eta)\, h(y) + \frac{\pat \Gt}{\Gt} \,
  \bigl( g(y) -h(y)\bigr) \right]_{y=\frac{\Gt}{Z_k k^2}}, \label{3.5}
\end{eqnarray}
where we abbreviated
\begin{equation}
h(y):=\frac{- y\, r'(y)}{1+r(y)}, \quad
g(y):=\frac{r(y)}{1+r(y)}.\label{hg}
\end{equation}
In Eq.\re{3.5}, we also defined the anomalous dimension
\begin{equation}
\eta:=-\pat \ln Z_k =-\frac{1}{Z_k}\, \pat Z_k, \label{3.6}
\end{equation}
which is matrix valued in field space similarly to $Z_k$; different
field variables can acquire different anomalous dimensions.  We would
like to draw attention to the appearance of the term $\sim \pat \Gt$
on the right-hand side of the flow equation. This term arises from
writing $\Gt$ into the argument of the cutoff function. It reflects
the fact that the cutoff adjusts itself under the flow of the spectrum
of $\Gt$.\footnote{Although $\Gt$ was also used as the argument of the
  cutoff in \cite{Reuter:1997gx}, the term $\sim \pat\Gt$ has been
  neglected in that calculation. The necessity of this term was
  pointed out to us by D.F.~Litim.}  Now it is useful to introduce (at
least formally) the Laplace transforms $\hi$ and $\gt$ of the
functions $h(y)$ and $g(y)$:
\begin{equation}
h(y)=\int\limits_0^\infty ds\,\hi\, \E^{-ys},\quad
g(y)=\int\limits_0^\infty ds\,\gt\, \E^{-ys}.\label{LThg}
\end{equation}
These Laplace transforms $\hi$ and $\gt$ can be viewed as cutoff
functions in Laplace space: they should drop off sufficiently fast for
large $s$ (small $s$) in order to regularize the infrared
(ultraviolet). For instance, the infrared constraint\re{3.4}
translates into 
\begin{equation}
h(0)=\int\limits_0^\infty ds\, \hi=1=
\int\limits_0^\infty ds\, \gt=g(0). \label{3.7}
\end{equation}
Additional useful identities for these functions are discussed in
Appendix \ref{cutoff}. Furthermore, introducing the functions $\Hl$ and
$\Gl$ by 
\begin{eqnarray}
\frac{d}{ds}\,\Hl=\hi,\quad \widetilde{H}(0)=0, \nonumber\\
\frac{d}{ds}\,\Gl=\gt,\quad \widetilde{G}(0)=0, \label{3.8}
\end{eqnarray}
a convenient form of the flow equation can be found which reads:
\begin{eqnarray}
\pat\Gk&=&\frac{1}{2} \int\limits_0^\infty \frac{ds}{s}
  \big(\Hl-\Gl\big) \, \pat\, \STr\, \exp\left( -s \frac{\Gt}{Z_k k^2}
  \right) \nonumber\\
&&+\frac{1}{2}\int\limits_0^\infty ds\, \gt\, \STr\, (2-\eta)\,
  \exp\left( -s \frac{\Gt}{Z_k k^2} \right). \label{3.9}
\end{eqnarray}
The great advantage of this form is that the right-hand side of the
flow equation has been transformed into a propertime
representation.\footnote{This representation of the flow equation
  should not be confused with the so-called propertime RG
  \cite{Liao:1994fp}. The latter represents a RG flow equation that is
  derived by RG improving one-loop formulas in propertime
  representation, and has been used in a variety of studies
  \cite{Liao:1995nm}. However, a propertime flow is generally {\em
    not} exact, as was proved in \cite{Litim:2001hk},
  \cite{Litim:2002}: generic propertime flows can neither be mapped
  onto exact flows in a derivative expansion nor correctly reproduce
  perturbation theory. By contrast, our flow equation is derived from
  an exact RG flow equation and corresponds to the {\em generalized}
  propertime flow proposed in \cite{Litim:2002}. The essential
  difference to (standard) propertime flows is the inclusion of
  $\pat\Gt$ terms. In the present work, particularly these terms will
  be important and will not be neglected. In agreement with
  \cite{Litim:2002}, our findings therefore suggest that propertime
  flows may be improved towards exact RG flows by including the
  $\sim\pat\Gt$ terms systematically.} The (super-)trace calculation
reduces to a computation of a heat-kernel trace for which there are
powerful techniques available.

Within the actual present truncation (neglecting $\Ga$), 
\begin{equation}
\Gk[A,\bA]=\int_x W_k(\tet)+\Ggh[A,\bA]+\Ggf[{
  A},\bA],\label{3.10}
\end{equation}
$\Gt$ still has a complicated structure which inhibits obtaining
general and exact results for the heat-kernel trace. Fortunately, a
general solution is not necessary; we merely have to project the
right-hand side onto the truncation, implying that we need only the
dependence of the right-hand side on the invariant $\tet=\case{1}{4}
F_{\mu\nu}^a F_{\mu\nu}^a$; other invariants occurring in the
heat-kernel trace are of no importance in the present truncation. Now
the crucial observation is that the heat-kernel dependence on $\tet$
can be reconstructed by performing the computation for the special
field configuration of a covariant constant magnetic field (as it is
explicitly defined in Eq.\re{A1}). In addition to this, we can perform
the computation for $A=\bA$. For this field configuration, the flow
equation finally depends only on the field parameter $\tet=\case{1}{4}
F_{\mu\nu}^a F_{\mu\nu}^a\equiv \case{1}{2}B^2$, where $B$ denotes the
strength of the magnetic field; the latter is pseudo-abelian and
points into a single direction in color space, characterized by a
color unit vector $n^a$. Extracting this $B$ dependence of the heat
kernel allows us to reconstruct the flow of $W_k(\case{1}{2} B^2)$.

It should be stressed that considering a covariant constant background
field is nothing but a technical trick to project onto the truncation;
we do not at all assume that such a background represents the vacuum
configuration of Yang-Mills theory, as is the case, e.g., in the Savvidy
vacuum model \cite{Savvidy:1977as}.

This trick also allows us to decompose the operator $\Gt$ into
linearly independent pieces. In particular, the gauge-field
fluctuations can be classified into modes with generalized transversal
(T) and longitudinal (L) polarization with respect to the
magnetic-field direction in spacetime, and into parallel $\|$ and
perpendicular $\bot$ modes with respect to the field direction in
color space. Introducing the corresponding projectors $P_{\text{L,T}}$
  and $P_{\|,\bot}$ (explicitly defined in App.\re{appDecomp}), the
  operator $\Gt$ can be represented as
\begin{eqnarray}
\Gt&=&\Pt\Pb \, \big[W_k'\, \Dt\big] + \Pl \Pb\, \left[
  \frac{1}{\alpha}\, \Dt \right] \nonumber\\
&&+\Pt\Pp\, \big[W_k'\, (-\pd^2)+W_k''\, \mathsf{S}\big]
  +\Pl\Pp\, \left[ \frac{1}{\alpha}\, (-\pd^2)\right]\nonumber\\
&&+P_{\text{gh}}\, \bigl[-D^2], \label{3.11}
\end{eqnarray}
where $\Gt\equiv\Gt[A=\bA,\bA]$, and we drop the bars from now
on. Here we also defined the operators
\begin{equation}
(\Dt)_{\mu\nu}^{ab}=(-D^2 \delta_{\mu\nu} +2\I\gb F_{\mu\nu})^{ab},
\quad 
\mathsf{S}_{\mu\nu}=\Fa_{\mu\alpha}\Fa_{\beta\nu}
\pd^\alpha\pd^\beta, \quad \Fa_{\mu\nu}=n^aF^a_{\mu\nu}.\label{3.12}
\end{equation}
The formal symbol $P_{\text{gh}}$ in Eq.\re{3.11} projects onto the
ghost sector, and $W_k'\equiv \frac{d}{d\theta} W_k(\theta)$; for
details about this decomposition, see App.~\ref{appDecomp}.

At this point, we are free to choose different cutoff wave-function
renormalizations $Z_k$ for each of the linearly independent parts in
Eq.\re{3.11}. If we were solving the flow equation exactly, the final
result would be independent of this choice; however, for a truncated
flow, a clever choice can seriously improve the approximation. With
regard to the form of $R_k$ in Eq.\re{3.2}, it is obvious that the
$Z_k$'s control the precise position at which the scale $k$ cuts off
the infrared of the momentum spectrum. Since the latter is determined
by the Laplace-type operators $-\pd^2,-D^2,\Dt$ in Eq.\re{3.11}, we
can cut them off at $k^2$ by choosing
\begin{equation}
Z_{\text{ghost},k}=1,\quad
Z_{\text{L},k}=\frac{1}{\alpha}, \quad
Z_{\text{T},k}=W_k'(0)\equiv\ZFk, \label{3.13}
\end{equation}
for ghost, longitudinal, and transversal fluctuations,
respectively.\footnote{In agreement with our approximation of
  neglecting $\Ga$, we also do not distinguish between the
  wave-function renormalization constant of the classical field $A$
  and that of the background field $\bar{A}$; if $\Ga$ were taken into
  account, this distinction would have to be made
  \cite{Pawlowski:2002}. A similar problem would occur, if we relaxed
  the constraint\re{3.4} for the cutoff; by keeping track of different
  wave-function renormalization constants, such a problem was
  explicitly solved in \cite{Litim:2002ce}.} This choice guarantees
that the longitudinal and ghost modes are cut off at the same point,
providing for a necessary cancellation. As a side effect, the flow
becomes independent of the gauge-fixing parameter $\alpha$, so that we
can implicitly choose Landau gauge $\alpha\to 0$, which is known to be
a fixed point of the flow \cite{Ellwanger:1995qf},\cite{Litim:1998qi}.
Finally, the transversal cutoff wave-function renormalization is set
equal to the gauge-field wave function renormalization, which can be
read off at the weak-field limit, $\Gk[A]|_{\text{w.f.}}\simeq
W_k'(0)\, \theta\equiv\frac{\ZFk}{4} F_{\mu\nu}^a F_{\mu\nu}^a$.

Using trace identities found in \cite{Reuter:1997gx}, the heat-kernel
trace occurring in Eq.\re{3.9} can be further reduced to
\begin{eqnarray}
\STr\, \E^{-s\frac{\Gt}{Z_k k^2}} &=& 
  \Tr_{x\text{L}} \E^{-\frac{s}{k^2} \left( \frac{W_k'}{\ZFk}(-\pd^2) +
      \frac{ W_k''}{\ZFk} \mathsf{S} \right)}
  -d\, \Tr_{x}\, \E^{-\frac{s}{k^2} \frac{W_k'}{\ZFk} (-\pd^2) }\nonumber\\
&&+\Tr_{x\text{cL}}\, \E^{-\frac{s}{k^2} \frac{W_k'}{\ZFk}\, \Dt}
  -\Tr_{x\text{c}}\, \E^{-\frac{s}{k^2} \frac{W_k'}{\ZFk}\, (-D^2)}
  -\Tr_{x\text{c}}\, \E^{-\frac{s}{k^2}\, (-D^2)},\label{3.14}
\end{eqnarray}
where the traces can act on spacetime ($x$), color (c), or Lorentz (L)
indices.  For the trace in Eq.\re{3.11} involving $\eta$
(matrix-valued), all terms in Eq.\re{3.14} containing $\ZFk$ will
acquire an anomalous-dimension contribution which we will also call
$\eta$ for simplicity:
\begin{equation}
\eta=-\pat \ln \ZFk=-\frac{1}{\ZFk}\, \pat \ZFk.\label{3.15}
\end{equation}
The various heat-kernel traces are computed in Appendix
\ref{heatkernel}. In order to display the result concisely, let us
define the auxiliary functions
\begin{eqnarray}
f_1(u)&=&\frac{1}{u^{d/2}}\left(\frac{(d-1)}{2}\, \frac{u}{\sinh{u}} +
  2\, u\, \sinh u \right), \nonumber\\
f_2(u)&=&\frac{1}{2} \frac{1}{u^{d/2}}\, \frac{u}{\sinh u},
  \label{3.16}\\
f_3(v_1,v_2)&=&\frac{1}{v_1^{d/2}}\, (1-v_2).%,
 \nonumber
\end{eqnarray}
Equipped with these abbreviations, the flow equation can be written as
\begin{eqnarray}
\pat W_k(\theta)&=&\frac{1}{2(4\pi)^{d/2}} \int\limits_0^\infty ds \left\{
  \gt\, \left[%\ftot\left(\frac{s}{k^2}, \frac{W_k}{\ZFk},B,\eta\right)
\sum_{l=1}^{N^2-1}\left( 
   2 (2-\eta)f_1\left(\frac{s}{k^2}\frac{W_k'}{\ZFk}\Bl\right) 
  -4 f_2\left(\frac{s}{k^2}\Bl\right) \right)\Bl^{d/2}
  \right.\right. \nonumber\\
&& \qquad\qquad\qquad\quad\qquad\,\,\,\,\left. 
  -(2-\eta)\, f_3\left(\frac{s}{k^2} \frac{W_k'}{\ZFk},
  \frac{W_k'}{W_k'+B^2  W_k''}\right)\right]
  \label{3.18}\\
&&\qquad\qquad\qquad\,\,\,\,\left.
  +\frac{1}{2s} \bigl[\Hl-\Gl \bigr]\, \pat\, %\ftot\left(\frac{s}{k^2},
%  \frac{W_k}{\ZFk},B,0\right)\right\},
\left[4 \sum_{l=1}^{N^2-1}\left(
 f_1-f_2\right)\Bl^{d/2}-2f_3\right]\right\},\nonumber
\end{eqnarray}
where $\Bl=\gb|\nu_l|B$, $\gb$ denotes the bare coupling, and $\nu_l$
represents the $l=1,\dots,N^2-1$ eigenvalues of the color matrix
$(n^aT^a)^{bc}$. The auxiliary functions $f_i$ in the last line are
understood to have the same arguments as in the first lines. It is
convenient to express the flow equation in terms of dimensionless
renormalized quantities,
\begin{eqnarray}
g^2&=&k^{d-4}\, \ZFk^{-1}\, \gb^2, \nonumber\\
\te&=&g^2\, k^{-d}\, \ZFk\, \theta\equiv k^{-4}\, \gb^2\, \theta,
  \label{3.19}\\
w_k(\te)&=&g^2\, k^{-d}\, W_k(\theta)\equiv k^{-4}\, \ZFk^{-1}\,
\gb^2\, W_k(k^4\te/\gb^2), \nonumber
\end{eqnarray}
and evaluate the derivative $\pat$ from now on at fixed $\te$ instead
of fixed $\theta$. As a result, the flow equation\re{3.18} turns into
\begin{eqnarray}
&&\text{}\!\!\!\!\!\!\!\pat w_k(\te)\nonumber\\
&&=-(4-\eta)\, w_k+4\,\te\, \wdot(\te) \nonumber\\
&&\quad+\frac{g^2}{2(4\pi)^{d/2}} 
\Bigg\{\!\int\limits_0^\infty \!ds\,
  \hi \left[4\!\sum_{l=1}^{N^2-1} 
    \Bigl(f_1(s\wdot b_l) \!-f_2(s b_l)\Bigr)b_l^{d/2}
    -2f_3\!\left(\!s\wdot,\frac{\wdot}{\wdot\!+2\te\wddot}\right)\!\right]
  \nonumber\\
&&\qquad\qquad\qquad-\eta\int\limits_0^\infty ds\,
  \gt \left[2 \sum_{l=1}^{N^2-1} f_1(s\wdot b_l)\,b_l^{d/2}
    -f_3\left(s\wdot,\frac{\wdot}{\wdot+2\te\wddot}\right)\right]
\label{3.20}\\
&&\qquad\qquad\qquad+\int\limits_0^\infty ds\bigl(\hi-\gt\bigr) \left[
  \left( \frac{\pat\wdot}{\wdot}-\frac{4\te\wddot}{\wdot}\right)
  2\sum_{l=1}^{N^2-1} f_1(s\wdot b_l)\,b_l^{d/2} \right.\nonumber\\
&&\qquad\qquad\qquad\qquad\quad\qquad\qquad\qquad\,\,\left.
    -\frac{2}{d} (\pat-4\te\partial_\te)
  \,f_3\left(s\wdot,\frac{\wdot}{\wdot+2\te\wddot}\right)
  \right]
\Bigg\},\nonumber
\end{eqnarray}
where $\wdot(\te)=\pd_\te w_k(\te)$, and we abbreviated
$b_l=|\nu_l|\sqrt{2\te}$. Equation\re{3.20} represents one of the main
results of the present work. Within the chosen truncation, this flow
equation leads to the full quantum effective action of Yang-Mills
theory upon integration from its initial condition at $\Lambda$ down
to $k=0$.

As a first comment, we would like to mention that we rediscover
the flow equation of \cite{Reuter:1997gx} if we perform an
expansion for weak magnetic field and if we neglect all terms
proportional to $\pat\Gt$. In order to isolate the latter from the
rest of Eq.\re{3.20}, the single factor of $\gt$ in the second line
should be represented as $\hi-(\hi-\gt)$, and then all terms
proportional to $(\hi-\gt)$ should be dropped (cf. Eq.\re{3.5}).

Obviously, the $\pat\Gt$ terms $\sim (\hi-\gt)$ modify the flow
equation extensively.\footnote{Incidentally, it is easy to show that
  no admissible cutoff shape function $r(y)$ exists such that
  $h(y)=g(y)$. Hence, the $\pat\Gt$ terms are present for all cutoff
  shape functions.} They seriously increase the degree of complexity
of this partial differential equation, so that neither an analytic nor
a numeric evaluation is straightforward. The next section will be
devoted to a search for the simplest possible and consistent
approximation.

Finally, we remark that the flow equation contains a seeming
divergence: in the limit of small $k$, the $s$ integrand may not be
bounded for $s\to\infty$, owing to the last term $\sim \sinh u$ in the
auxiliary function $f_1$ given in Eq.\re{3.16}. However, this
divergence is well understood and can be controlled. It arises from
the Nielsen-Olesen unstable mode \cite{Nielsen:1978rm} in the operator
$\Dt$, and can be traced back to the fact that the gluon-spin coupling
to the constant magnetic field can lower its energy below zero.
Because of this mode, the covariant constant magnetic field is known
to be unstable, if considered as the quantum vacuum state of
Yang-Mills theory. The divergence can be identified as a pole at
complex infinity. The $s$ integral can be properly defined by analytic
continuation, resulting in a real part as well as an imaginary part.
The real part is indeed important because it contributes to the
$\beta$ function and the form of the effective action in a
perturbative computation (see below). The imaginary part is
interpreted as a measure for the instability of the constant-field
vacuum.

As we have stressed before, the constant-magnetic-field background is
just a calculational tool in the present context, and the validity of
the flow equation is not based on this background. Therefore, the $s$
integral can be properly defined by analytic continuation around this
pole at complex infinity. The resulting real part will be a valid and
important contribution to the flow, but the imaginary part is of no
relevance here. If we were really interested in a constant-field
vacuum, the flow generated by this imaginary part would describe how
the instability develops upon integrating out the unstable mode in a
Wilsonian sense.

\section{Running gauge coupling in $d=4$}
\label{expansion}

In order to find a strategy for solving the flow equation\re{3.20}
within a first simple approximation, let us take a closer look at the
standard procedure employed for ordinary cutoffs without $\pat \Gt$
terms. In such a case, the partial differential equation can be
rewritten as an infinite set of coupled ordinary first-order
differential equations by expanding the truncation, e.g.,
\begin{equation}
w_k(\te)=\sum_{i=1}^\infty \frac{w_i}{i!}\, \te^i, \quad w_1=1.
  \label{4.1} 
\end{equation}
Note that, owing to the choice\re{3.13} for $Z_k$ and the
definition\re{3.19}, $w_1=1$ is fixed, so that the generalized
coupling $W_1$ is traded for the anomalous dimension $\eta$. As a
result, we obtain infinitely many flow equations for the couplings
$w_i$ which, for an ordinary cutoff, read
\begin{equation}
\pat w_i\big|_{\text{ordinary cutoff}}
=X_i(\eta,w_2,\dots,w_{i+1}), \quad i=2,3,\dots\,\,. \label{4.2}
\end{equation}
Equation\re{4.2} is supplemented by an additional equation for $\eta$.
The functions $X_i$ are obtained as the $i$th coefficient of the $\te$
series expansion of the flow equation's right-hand side.  This
infinite tower of equations is then approximated by a finite one by
setting all $w_i=0$ by hand for some $i>i_{\text{trunc}}$, resulting
in $i_{\text{trunc}}$ equations for $i_{\text{trunc}}$ variables.  The
quality of this further truncation can be checked by varying
$i_{\text{trunc}}$.

This recipe cannot be directly applied to the present case involving
the spectrally adjusted cutoff because an expansion of the flow
equation\re{3.20} will be of the form
\begin{equation}
\pat w_i
=X_i(\eta,w_2,\dots,w_{i+1})+Y_i(\eta,w_2,\dots,w_{i+1}; \pat w_2,
\dots, \pat w_{i+1}), \quad i=2,3,\dots\,\,. \label{4.3}
\end{equation}
It is tempting to truncate this tower by setting not only
$w_{i>i_{\text{trunc}}}=0$ by hand, but also $\pat
w_{i>i_{\text{trunc}}}=0$. This is too naive, however, because
all $\pat w_i$, if understood as the left-hand side of Eq.\re{4.3},
receive nonzero contributions on the right-hand side, even if
${i>i_{\text{trunc}}}$. Neglecting these right-hand sides would
correspond to neglecting some $w_i$'s which are in the truncation 
${i\leq i_{\text{trunc}}}$.

In order to apply the above-mentioned recipe, we have to bring
Eq.\re{4.3} into the form of Eq.\re{4.2}, i.e., we have to solve for
the $\pat w_i$'s. Formally, this is possible by observing that the
functions $Y_i$, as they are derived from Eq.\re{3.20}, are linear in
all $\pat w_i$ and $\eta$, and the $X_i$ are also linear in $\eta$.
Introducing a ``vector'' $\vec{w}_t$ with components
\begin{equation}
\vec{w}_t:=\Bigl\{\begin{array}{ll}
                     w_{t\,1}=-\eta\\
                     w_{t\,i}=\pat w_i
                     \,\,\,\text{for}\,\,\,i=2,3,\dots
                  \end{array}\Bigr\}, \label{4.4} 
\end{equation}
equation\re{4.3} can be written as\footnote{The meaning of the
  quantities $X_i$ and $Y_{ij}$ changes here slightly, because the
  $\eta$ and $\pat w_i$ dependence is pulled out compared to
  Eq.\re{4.3}.} 
\begin{equation}
w_{t\,i}=X_i(w_2,\dots,w_{i+1})+Y_{ij}(w_2,\dots,w_{i+1})\,
w_{t\,j},\label{4.5}
\end{equation}
or symbolically, $\vec{w}_t=\vec{X}+Y\cdot\vec{w}_t$. Provided that the
operator $1-Y$ is invertible, the desired solution is formally given
by 
\begin{equation}
\vec{w}_t=\frac{1}{1-Y}\cdot \vec{X}, \label{4.6}
\end{equation}
where the right-hand side is a function of $w_2,w_3,\dots$ only. Now,
the approximation strategy for the ordinary cutoff can be applied to
Eq.\re{4.6}. Nevertheless, the resulting finite tower of differential
equations is substantially different from the ordinary case, even for
the smallest $i_{\text{trunc}}$. This is because $X_i$ and $Y_{ij}$
are generally nonzero, even for $i,j>i_{\text{trunc}}$, since they
depend on the remaining $w_{i\leq i_{\text{trunc}}}$ (and numbers such
as $d$ and $N$). And since they are infinite dimensional, we find an
infinite number of terms on the right-hand side of the flow equations,
in contrast to a finite number for ordinary cutoffs. 

For the remainder of this section, we shall evaluate Eq.\re{4.6} in
the simplest possible way by neglecting all $w_i$'s with $i=2,3,\dots$
and retaining only the anomalous dimension $\eta$, which is related to
the $\beta$ function of Yang-Mills theory via
\begin{equation}
\beta(g^2)\equiv \pat g^2 = (d-4+\eta)\, g^2, \label{4.7}
\end{equation}
so that in $d=4$ we simply have $\beta(g^2)=\eta\, g^2$. We would like
to stress that the approximation of neglecting all $w_i$'s at this
stage is not at all equal to neglecting them right from the beginning.
This further truncation is only consistent {\em after} we have
disentangled the flows of all $w_i$'s by virtue of Eq.\re{4.6}.

For the investigation of the $\eta$ equation, corresponding to the
first component of the vector equation\re{4.6}, it is useful to scale
out the coupling constant, so that $\vec{X}$ and $Y$ no
longer depend on the coupling:
\begin{equation}
\vec{X}\to G\,\vec{X}, \quad Y\to G\, Y,\quad
G:=\frac{g^2}{2(4\pi)^{d/2}}. \label{4.8}
\end{equation}
In $d=4$, the convenient coupling $G$ is related to the standard
strong coupling constant $\alpha_{\text{s}}\equiv \frac{g^2}{4\pi}
=8\pi G$. Using Eq.\re{4.8}, we can perform a ``perturbative''
expansion of the $\eta$ equation:
\begin{eqnarray}
-\eta&\equiv& w_{t\, 1}=\left(\frac{1}{1-G\, Y}\right)_{1j}\, G\,X_j
=G(1+GY+G^2Y^2+\dots)_{1j}\,X_j \nonumber\\
&=&G\left(\sum_{m=0}^\infty G^m\, Y^m\right)_{1j} X_j. \label{4.9}
\end{eqnarray}
The explicit representation of $Y$ and $\vec{X}$ can be found by
inserting the expansions developed in Appendix \ref{series} into
Eq.\re{3.20}, and performing the propertime $s$ integration; the
latter results in the moments $h_j,g_j$ of the cutoff functions $\hi,\gt$,
\begin{equation}
h_j:=\int\limits_0^\infty ds\, s^j\, \hi,\quad
g_j:=\int\limits_0^\infty ds\, s^j\, \gt, \label{4.10}
\end{equation}
which are discussed in Appendix \ref{cutoff}. In conclusion, we find:
\begin{eqnarray}
X_i&=&-2^{i+1}\, \tau_i\, h_{2i-d/2}\, i! \left((d-2)
  \frac{(2^{2i}-2)}{(2i)!} \, B_{2i} -\frac{4}{(2i-1)!}\right),
  \nonumber\\
Y_{ij}&=&A_{ij}+B_{ij}+C_{ij}, \label{4.11}
\end{eqnarray}
where $B_{2i}$ denotes the Bernoulli numbers, and the auxiliary
matrices $A,B,C$ are given by ($i,j=1,2,\dots$):
\begin{eqnarray}
A&=&\left\{\begin{array}{l}
A_{i1}=0 \\
            A_{ij}=0\,\,\,\text{if}\,\,\, j>i+1 \\
            A_{ij}=\frac{i!}{(j-1)!} \left[2^n \tau_n
              (h_{2n-d/2}-g_{2n-d/2}) \left((d-1)
                \frac{2^{2n}-2}{(2n)!}B_{2n}-\frac{4}{(2n-1)!} 
                  \right)\right]_{n=1+i-j}
         \end{array} \right.\nonumber\\
B&=&\left\{\begin{array}{l}
            B_{ij}=0\,\,\,\text{if}\,\,\, j>1\\
            B_{i1}=-2^i\tau_i h_{2i-d/2} i! \left((d-1)\frac{
                   2^{2i}-2}{(2i)!}B_{2i}-\frac{4}{(2i-1)!}\right)
          \end{array} \right.\nonumber\\
C&=&\left\{\begin{array}{l}
            C_{ij}=0\,\,\,\text{if}\,\,\, j\neq i+1\\
            C_{i,i+1}=-\frac{4}{d}\,i\,(h_{-2}-g_{-2})
          \end{array} \right. .\label{4.12}
\end{eqnarray}
These explicit representations\re{4.11} and\re{4.12} can be inserted
into Eq.\re{4.9}, and the anomalous dimension and the $\beta$ function
can be computed straightforwardly to any finite order in perturbation
theory within our truncation. 

As an example, let us compute the two-loop $\beta$ function in $d=4$
spacetime dimensions for SU($N$) gauge theories:
\begin{equation}
\beta(g^2)=\pat g^2=-\frac{22N}{3} \frac{g^4}{(4\pi)^2} 
  -\left( \frac{77N^2}{3}- \frac{127(3N^2-2)}{45}\bigl(
    h_{-2}-g_{-2}\bigr) h_2\, \tau_2 
    \right)\frac{g^6}{(4\pi)^4}+\dots\,\,. \label{4.13}
\end{equation}
Here we already used the fact that $h_0=1=g_0$ are independent of the
shape of the cutoff function, so that the one-loop coefficient turns
out to be universal as it should be and agrees with the standard
perturbative result; this should serve as a (rather trivial) check of
our computation. Within our truncation, the two-loop coefficient does
depend on the cutoff function. In order to compare with
\cite{Reuter:1997gx} where the $\pat \Gt$ terms have been neglected,
we choose the exponential cutoff defined in Eq.\re{D4}, implying that
$g_{-2}=1$, $h_{-2}=2\zeta(3)\simeq 2.404$, and $h_2=1/6$ as computed
in Appendix \ref{cutoff}. From Appendix \ref{SU3}, we take over that
$\tau_2^{N=2}=2$ and $\tau_2^{N=3}=9/4$. Inserting all these numbers
and comparing this to the perturbative two-loop result,
\begin{equation}
\beta_{\text{pert.}}(g^2)=-\frac{22N}{3}\, \frac{g^4}{(4\pi)^2}
-\frac{68N^2}{3}\, \frac{g^6}{(4\pi)^4}+\dots, \label{4.14}
\end{equation}
we find a remarkable agreement of 99\% for the two-loop coefficient
for SU(2), and 95\% for SU(3). This should be compared to only a 113\%
agreement of these coefficients in the case where the $\pat\Gt$ terms
are neglected \cite{Reuter:1997gx}. The inclusion of these terms
appears to represent a serious improvement. 

However, the picture is not so rosy as it seems to be in view of this
result. The reason is that our two-loop result is cutoff-scheme
dependent, and we may easily choose a cutoff with a worse agreement at
two loop.\footnote{Actually, this point is even more subtle
  \cite{Pawlowski:2002}: cutoff-scheme independence of the two-loop
  $\beta$ function coefficient holds only for mass-independent
  regulators. The regulator $R_k$ obviously does not belong to this
  class, so that cutoff-scheme dependence has to be expected. However,
  since we have no information about the true two-loop coefficient for
  our $k$-dependent regulator $R_k$, we shall still use the two-loop
  coefficient for a mass-independent scheme as our benchmark test.}
Only recently has it been explicitly shown how to obtain the correct
scheme-independent two-loop $\beta$ function within the framework of
the exact RG \cite{Pawlowski:2002}; for this, a careful distinction
has to be drawn between the running of the coupling with respect to
$k$ or the RG scale $\mu$ (see also \cite{Bonini:1997bk}). Within our
truncation here, we can nevertheless turn the argument around by
remarking that the exponential cutoff is obviously well suited for the
present truncation in the sense that it minimizes the combined effect
of the neglected terms such as $w_2$, $(F_{\mu\nu}^a
\widetilde{F}_{\mu\nu}^a)^2$, or $\Ga$, etc. on the two-loop
coefficient. This particularly justifies the use of the exponential
cutoff for an investigation of the complete sum in Eq.\re{4.9} and the
strong-coupling domain.

Let us summarize what has been achieved so far: in order to extract
the flow of the gauge coupling, the flow equation\re{3.20} has to be
studied near $\te=0$, where the information about $\eta$ is
encoded. This suggests an expansion of $w_k(\te)$ in powers of
$\te$, leading to completely disentangled flow equations\re{4.6}
for the generalized couplings $\eta,w_2,w_3,\dots$. However, since the
original flow equation\re{3.20} is represented as a parameter
integral, its expansion can be asymptotic, which implies that the
series expansion in terms of the coupling $G$ in Eq.\re{4.9} will be
asymptotic as well. This agrees with the general expectation that
perturbative expansions of quantum field theories are generically
asymptotic. 

In practice, this means that the coefficients (for later convenience,
we shift the index $m$ here)
\begin{equation}
a_m:=-(Y^{m-1})_{1j} X_j, \quad m=1,2,\dots \label{4.15}
\end{equation}
in Eq.\re{4.9} grow rapidly, so that any arbitrarily large but finite
truncation of the series does not make sense. It turns out that these
coefficients grow even more strongly than factorially and alternate in
sign for the exponential cutoff (we shall comment on other cutoffs
later). This does not mean that any physical meaning is lost, but,
loosely speaking, that we have expanded an integrand which we should
not have expanded. Yet, there are well-defined mathematical tools for
reconstructing the integrand representation out of the diverging sum
\cite{Hardy}.  In other words, we are looking for a (well-defined)
integral representation that upon asymptotic expansion leads to a
series that agrees with Eq.\re{4.9}. As is known also from various
physical examples \cite{Guillou}, just taking only the leading
growth of the coefficients into account leads to a good approximation
of the integral representation.
 
Concerning the coefficients $a_m$, the leading growth (l.g.) can be
isolated in the term that contains the highest component of $\vec{X}$,
i.e., $X_{m}$, yielding
\begin{equation}
a^{\text{l.g.}}_1=-X_1,\quad a^{\text{l.g.}}_2=-Y_{12}X_2,\quad
a^{\text{l.g.}}_m=-Y_{12}Y_{23}Y_{34}\dots Y_{m-1,m}
X_{m}. \label{4.16} 
\end{equation}
Inserting the representations\re{4.12} into Eq.\re{4.16} for the
exponential cutoff, we find
\begin{eqnarray}
a^{\text{l.g.}}_m=4(-2c)^{m-1}
\frac{\Gamma(m+3(N^2-1))\Gamma(m+1)}{\Gamma(3N^2-2)} \, \tau_m\, 
B_{2m-2} \left(2 \frac{2^{2m}-2}{(2m)!} B_{2m} -\frac{4}{\Gamma(2m)}
\right), \label{4.17}
\end{eqnarray}
where we abbreviated $c=2\zeta(3)-1$. Let us first concentrate on
SU(2), where $\tau_m=2$ for $m=1,2,\dots$ (see Appendix \ref{SU3});
let us nevertheless retain the $N$ dependence in all other terms in
order to facilitate the generalization to SU(3). 

Actually, Eq.\re{4.17} also contains subleading terms. First, we
observe that the last term $\sim 1/\Gamma(2m)$ is negligible compared
to the term $\sim B_{2m}$ for large $m$. Nevertheless we also retain
this subleading term, since the $m=1$ term contributes significantly
to the one-loop $\beta$-function coefficient which we want to maintain
in our approximation. Furthermore using the identity
\begin{equation}
B_{2m-2}=\frac{(-1)^m \Gamma(2m-1)}{2^{2m-3} \pi^{2m-2}} \,
\zeta(2m-2), \label{4.18}
\end{equation}
it is tempting to use the $\zeta$-function representation
\begin{equation}
\zeta(z)=\sum_{l=1}^\infty \frac{1}{l^z},\label{4.19}
\end{equation}
and retain only the $l=1$ term, since the others are subleading.
Whereas this approximation is indeed justified for the Bernoulli
number $B_{2m}$ in Eq.\re{4.17}, we have to retain the full $\zeta$
function for the $B_{2m-2}$ factor, since here we encounter the
$\zeta$ function at zero argument for $m=1$ where Eq.\re{4.19} is no
longer valid. In conclusion, we resum the complete coefficient
$a^{\text{l.g.}}_m$ as displayed in Eq.\re{4.17}, including the leading
and also subleading terms. 

In the spirit of Borel summation \cite{Hardy}, we introduce integral
representations for the special functions occurring in Eq.\re{4.17}, in
particular the representation \cite{GR}
\begin{equation}
\Gamma(2m-1)\, \zeta(2m-2)=\frac{1}{1-2^{3-2m}} \int\limits_0^\infty
dt\, \frac{\E^t}{(e^t+1)^2}\, t^{2m-2}, \quad m>1/2 \label{4.20}
\end{equation}
for the $\zeta$ function in Eq.\re{4.18}, and an Euler $B$ function
representation for a combination involving the last term in
Eq.\re{4.17}:
\begin{eqnarray}
\frac{\Gamma(m+3(N^2-1))\Gamma(m+1)}{\Gamma(2m)} &\equiv&
  \frac{\Gamma(m+3(N^2-1))}{\Gamma(m)}\,
  \frac{\Gamma(m+1)}{\Gamma(m)}\, B(m,m)\nonumber\\
&=&m(m+1)\dots (m+3N^2-4)\cdot m \int\limits_0^1 ds\, s^{m-1}\,
  (1-s)^{m-1} \nonumber\\
&=&\int\limits_0^1ds\left[ \left(\frac{d}{ds}\right)^{(3N^2-3)}
  s^{m+(3N^2-4)} \right] \left( \frac{d}{ds'} \, s'{}^m
  \right)_{s'=1-s}. \label{4.21}
\end{eqnarray}
For the remaining $\Gamma$ functions, we use the standard Euler
representation. Exploiting these identities, we are able to resum
Eq.\re{4.9} to this order:
\begin{eqnarray}
\eta&=&-G\left(\sum_{m=1}^\infty G^{m-1}\, Y^{m-1}\right)_{1j} X_j
  \stackrel{\text{l.g.}}{\simeq} \sum_{m=1}^{\infty} a_m^{\text{l.g.}}\,
  G^m\nonumber\\
&=:&\eta_{\text{a}}+\eta_{\text{b}}, \label{4.22}
\end{eqnarray}
where $\eta_{\text{a}}$ is related to the term $\sim B_{2m}$ in
Eq.\re{4.17}, whereas $\eta_{\text{b}}$ is related to the term $\sim
1/\Gamma(2m)$. The integral representation of $\eta_{\text{a}}$ reads
\begin{equation}
\eta_{\text{a}}^{N=2}\!
=\frac{32 N G}{\Gamma(3N^2\!-2)\pi^2} \sum_{l=1}^\infty\!
\int\limits_0^\infty\!%&& 
ds_1 ds_2 dt \frac{\E^{t-(s_1+s_2)}}{(\E^t+1)^2}
\, \frac{s_1 s_2^{3N^2-3}}{l^2} \! %\nonumber\\
%&&\times
\left[ S\!\left(\!\frac{cGs_1s_2t^2}{2\pi^4l^2} \right)
-\frac{1}{2}\, S\!\left(\!\frac{cGs_1s_2t^2}{8\pi^4l^2} \right)\!\right]\!,
\label{4.23}
\end{equation}
where we defined the sum
\begin{equation}
-\sum_{m=1}^\infty \frac{(-q)^{m-1}}{1-2^{3-2m}} 
= 1+\sum_{j=0}^\infty \frac{q}{2^j +\frac{q}{2^j}}
=:S(q). \label{4.24} 
\end{equation}
The first sum arises from the asymptotic expansion and is strictly
valid only for $|q|<1$; however, the second sum is valid for arbitrary
$q$, apart from simple poles at $q=-2^{2j}$, and rapidly converging, so
that this equation should be read from right to left.
 
The second part $\eta_{\text{b}}$ deserves a comment: as it arises
from the last term in Eq.\re{4.17}, $\sim 1/\Gamma(2m)$, it originates
in the last term $\sinh u$ of the auxiliary function $f_1$ in
Eq.\re{3.16}, which stems from the lower end of the spectrum; in
particular, it contains the Nielsen-Olesen unstable mode. This mode
is reflected in a simple pole in the following integral
representation for $\eta_{\text{b}}$. This pole gives rise to an
imaginary part of the full integrand. As we have stressed above, the
imaginary part created by this unstable mode is of no relevance for
the flow equation here, so that the proper treatment of the integral
results in a principal-value prescription maintaining the important
real part. For a numerical realization, this prescription can best be
established by rotating the $t$ integral arising from Eq.\re{4.20} by
an angle of, e.g., $\pi/4$ from the real axis into the upper complex
plane and then taking the real part. In conclusion, we get:
\begin{equation}
\eta_{\text{b}}^{N=2}\!=-\frac{32 N G}{\Gamma(3N^2\!-\!2)} \, \text{Re}
  \int\limits_0^1\!\! ds \!\int\limits_0^\infty\!\!
  \case{(1\!+\I)}{\sqrt{2}}dt 
  \frac{\E^{\frac{1+\I}{\sqrt{2}} t}}{(\E^{\frac{1+\I}{\sqrt{2}}
  t}\!+1)^2} \!\left(\!\frac{d}{ds}\!\right)^{\text{}\!\!\!(3N^2-3)}\!\!
  \frac{d}{ds'} \, 
  s^{3N^2-3} s'\, 
%  \nonumber\\
%&&
 S\!\left(\! -\I \frac{cG s s' t^2}{2\pi^2}
  \right)\!\!\bigg|_{s'=s-1}\!\!. \label{4.25}
\end{equation}
Although it seems that we have seriously complicated the problem by
trading the single $m$ sum in Eq.\re{4.9} for a number of integrals
and sums, we stress that all integrals and sums in Eqs.\re{4.23}
and\re{4.25} are finite and well defined. 

Before we present numerical evaluations of these integrals and sums,
let us discuss some features analytically. For small coupling
$G=g^2/[2(4\pi)^2]$, we can again expand the integrals asymptotically
and obtain
\begin{equation}
\eta_{\text{a}}=\frac{2}{3} N\, \frac{g^2}{(4\pi)^2}+\dots, \quad
\eta_{\text{b}} =- 8 N\, \frac{g^2}{(4\pi)^2}+\dots, \label{4.26}
\end{equation}
so that we rediscover the one-loop $\beta$ function (cf. Eq.\re{4.13})
as a check. Next, we observe that $\eta_{\text{a}}$ (containing the
true leading-order growth of the $a_m^{\text{l.g.}}$'s) is positive
not only for small but arbitrary $G$. In order to extract large-$G$
information, we note that the sum $S(q)$ can be fitted by 
\begin{equation}
S(q)\simeq c_1 \sqrt{ \frac{1}{c_1^2}-c_2\sqrt{q} +q}, \quad
  c_1\simeq2.27, \quad c_2\simeq 0.7 \label{4.27}
\end{equation}
within 1\% accuracy, implying that $S(q)\simeq c_1\sqrt{q}$ for large
$q$. In this limit, which corresponds to large $G$, $\eta_{\text{a}}$
can be evaluated analytically and we find
\begin{equation}
\eta_{\text{a}}^{N=2}(G\gg1)\simeq \frac{24Nc_1}{\pi^4}
\sqrt{\frac{c}{2}} \zeta(3)\ln 2\,\Gamma(5/2)
\frac{\Gamma(3N^2-3/2)}{\Gamma(3N^2-2)}\, G^{3/2} \simeq 3.24 \,
G^{3/2}.\label{4.28}
\end{equation}
Without going into details, we note that there exist fits for $S(-\I
q)$ similarly to Eq.\re{4.27} involving a square-root behavior for
large $q$ (large $G$). It turns out that the $G^{3/2}$ coefficient
vanishes exactly, so that\footnote{We do not evaluate the precise
  coefficient of the $G^{1/2}$ here, since the large-$G$ expansion
  introduces artificial singularities for the $t$ integration at $t\to
  0$. A more careful treatment reveals that the coefficient is
  positive.}
\begin{equation}
\eta_{\text{b}}^{N=2}\sim +G^{1/2}. \label{4.29}
\end{equation}
Obviously, $\eta_{\text{b}}$ is subleading for large $G$, which agrees
with the fact that it arises from subleading parts in the coefficients
$a_m^{\text{l.g.}}$. Moreover, $\eta_{\text{b}}$ becomes positive for
large $G$, so that there should be a zero in between. This is already
the first sign of an infrared stable fixed point at which
$\eta(G_\ast)=0$.

\begin{figure}
\begin{center}
\begin{picture}(120,57)
\put(-8,-10){
\epsfig{figure=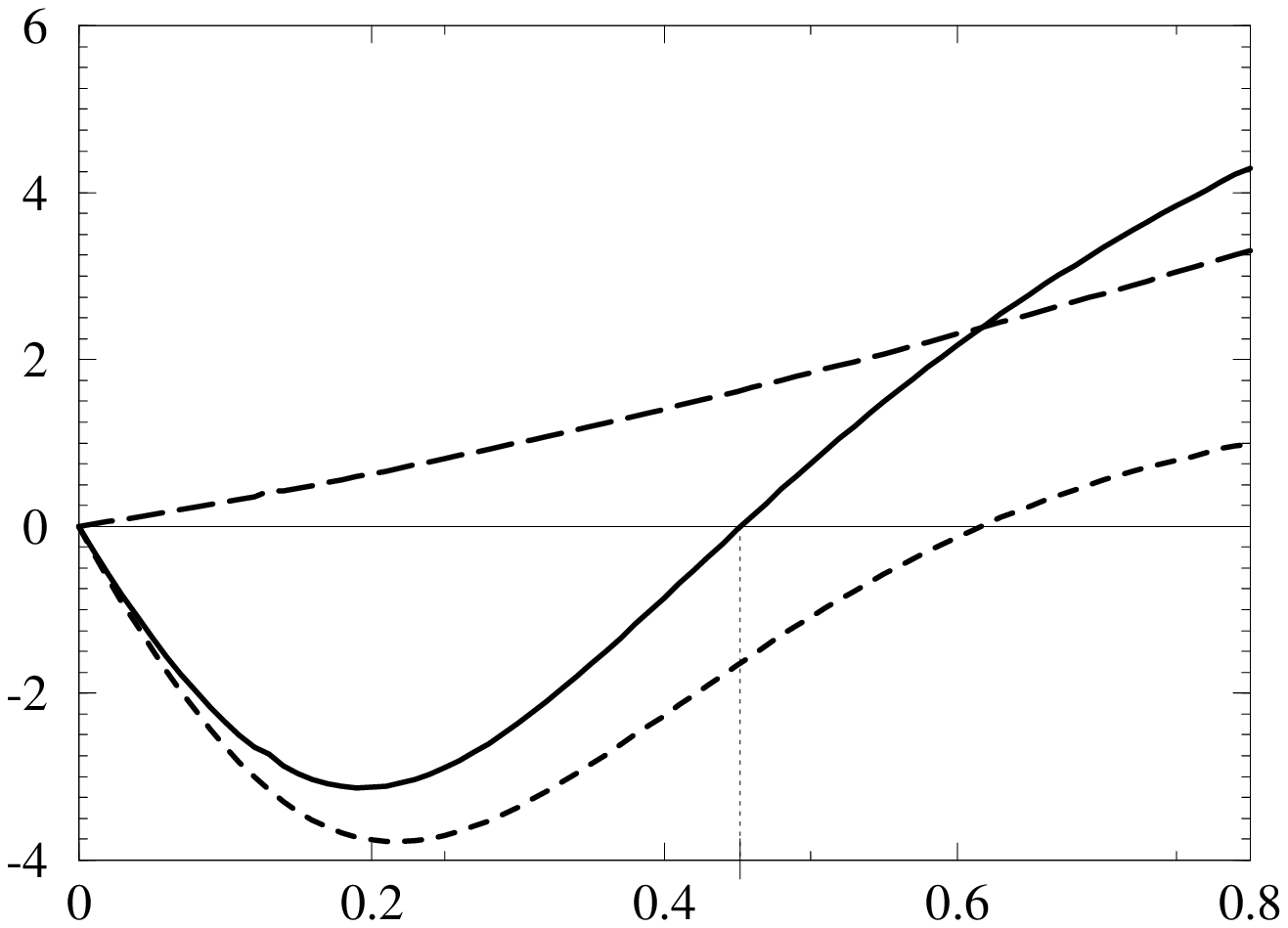,width=14cm}}
\put(5,61){$\eta$}
\put(25,59){SU(2)} 
\put(92,-5){$G$} 
\put(98,31){$\eta_b$}
\put(98,45){$\eta_a$}
\put(98,57){$\eta$} 
\put(66,-3){$G_\ast^{N=2}$}
\end{picture} 
\end{center} 
\caption{Anomalous dimension $\eta=\beta(g^2)/g^2$ for SU(2) Yang-Mills
  theory in $d=4$ versus $G=g^2/[2(4\pi)^2]$. The long-dashed line
  represents the contribution $\eta_a$, the short-dashed line
  $\eta_b$, as defined in Eqs.\re{4.23} and\re{4.25}; the solid line
  is the sum of both.}
\label{figSU2} 
\end{figure}

For a numerical evaluation of $\eta_{\text{a}}$ and $\eta_{\text{b}}$,
we employ the representations given in Appendix \ref{numerics}. We
depict the anomalous dimension $\eta$ and its parts $\eta_{\text{a}}$
and $\eta_{\text{b}}$ in Fig.~\ref{figSU2} for the gauge group
SU(2). The plots agree with the analytical estimates given above, and
we find an infrared stable fixed point at 
\begin{equation}
G_\ast^{N=2}\simeq 0.45 \quad \Rightarrow\quad 
\alpha_\ast^{N=2}\simeq 11.3. \label{4.30}
\end{equation}
By virtue of Eq.\re{4.7}, the running gauge coupling approaches this
fixed point upon lowering the scale $k$ in the infrared, implying
scale invariance. The complete flow of the coupling is obtained by
integrating $\beta(g^2)\equiv \pat g^2=\eta\, g^2$ and has been plotted
already in Fig.~\ref{figalpha}.

\begin{figure}
\begin{center}
\begin{picture}(120,57)
\put(-8,-10){
\epsfig{figure=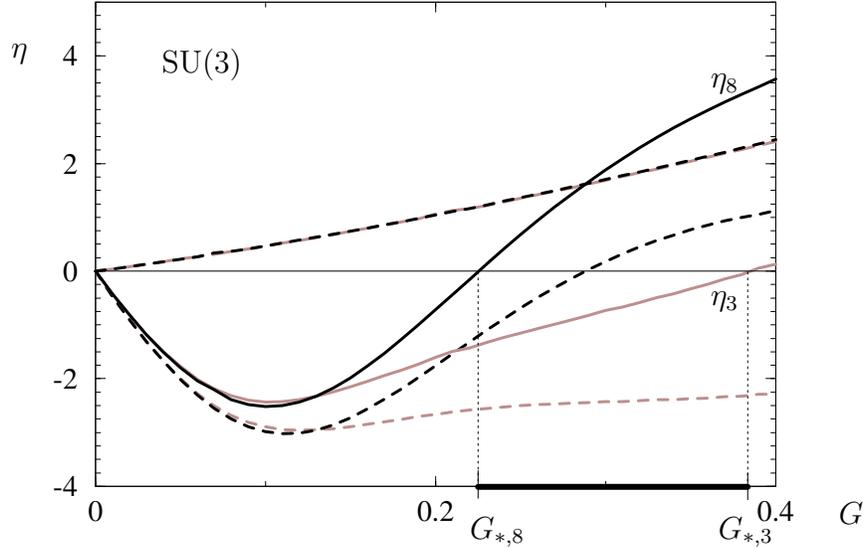,width=14cm}}
\put(5,61){$\eta$}
\put(25,59){SU(3)} 
\put(115,-1){$G$} 
%\put(98,31){$\eta_b$}
%\put(98,45){$\eta_a$}
\put(98,57){$\eta_8$} 
\put(98,28){$\eta_3$} 
\put(66,-3){$G_{\ast,8}$}
\put(99,-3){$G_{\ast,3}$}
\end{picture} 
\end{center} 
\caption{Anomalous dimension $\eta=\beta(g^2)/g^2$ for SU(3) Yang-Mills
  theory in $d=4$ versus $G=g^2/[2(4\pi)^2]$. The black lines
  correspond to $\eta_8^{N=3}$, the grey lines to $\eta_3^{N=3}$, as
  defined in Eq.\re{4.31}. The meaning of the dashed lines is as in
  Fig.~\ref{figSU2}.}
\label{figSU3} 
\end{figure}

For the gauge group SU(3), we do not have the explicit representation
of the color factors $\tau_m$ at our disposal. As discussed in
Appendix \ref{SU3}, we instead study the two extremal cases for the
color vector $n^a$ pointing into the 3 or 8 direction in color
space. Inserting the corresponding quantities $\tau^{N=3}_{i,3}$ or
$\tau^{N=3}_{i,8}$ as found in Eq.\re{C5} into Eq.\re{4.17} allows us
to display the anomalous dimension $\eta^{N=3}$ in terms of the
formulas deduced for SU(2):
\begin{eqnarray}
\eta^{N=3}_3&=&\frac{2}{3}\, \eta^{N=2}\Big|_{N\to 3} + \frac{1}{3}
\eta^{N=2} \Big|_{N\to 3,c\to c/4}, \nonumber\\
\eta^{N=3}_8&=&\eta^{N=2}\Big|_{N\to 3,c\to 3c/4}. \label{4.31}
\end{eqnarray}
The notation here indicates that the quantities $N$ and
$c=2\zeta(3)-1$ appearing on the right-hand sides of Eqs.\re{4.23}
and\re{4.25} will be replaced in the prescribed way. Figure
\ref{figSU3} depicts our numerical results, and we identify the position
of the infrared fixed point in the interval
\begin{equation}
G_\ast^{N=3}=[G_{\ast,8},G_{\ast,3}]\simeq[0.225,0.385]\quad
\Rightarrow \quad
\alpha_\ast^{N=3}\simeq[5.7,9.7]. \label{4.32}
\end{equation}
This uncertainty of the precise position of the fixed point is not a
shortcoming of the techniques involved (e.g., using the
covariant-constant magnetic background), but is due to our ignorance
of the exact color factors $\tau_m$.
 
Let us conclude this section with some remarks on the resummation:
first, we should stress that the results for the fixed point are
derived from a resummation of leading and subleading parts of the
complete asymptotic series\re{4.9}. We have checked that the
sub-subleading parts (not included in the present resummation)
alternate in sign, so that their contribution will be regular.
However, we were not able to systematize the sub-subleading terms in a
way that they can be resummed further in a consistent way.

Secondly, we performed the computation for the exponential cutoff. It
would be desirable to test the stability of the fixed point by using
different cutoffs. Unfortunately, we could not find another cutoff
shape function $r(y)$ for which the resummation could be done. For
many cutoff shape functions used in the literature, the series is also
asymptotic and alternating, but the sign changes not from one
coefficient to the other but from one group of coefficients to the
next, i.e., $a_{n_1}\dots a_{n_2}>0$ and $a_{n_2+1}\dots a_{n_3}<0$
for $n_1<n_2<n_3$, etc. The outstanding role of the exponential cutoff
may be attributed to its close relation to the Bernoulli numbers and
their properties.

Let us finally stress once more that the generalized Borel resummation
of Eq.\re{4.9} does not represent an uncontrolled extrapolation of
finite-order perturbation theory. As we have the exact all-order
result at our disposal, the resummation corresponds simply to an --
also mathematically -- well-defined transformation of a series into an
integral.

\section{The role of the spectrally adjusted cutoff}
\label{role}

This short section is devoted to a heuristic discussion of the special
role played by the spectrally adjusted cutoff in this work, focusing on
the truncation employed. 

The spectral adjustment of the cutoff function to the spectral
flow of $\Gt$ arises from two sources: first, from using $\Gt$ in the
argument of $R_k$ and, second, from including a carefully chosen
wave-function renormalization constant $Z_k$ in the cutoff. The latter
technique is well known in the problem of calculating anomalous
dimensions in scalar and fermionic theories. In order to get a
feeling for these two improvements, let us first consider the flow
equation, neglecting all $\pat\Gt$ and $\pat Z_k$ terms. For the
anomalous dimension, we would then obtain:
\begin{equation}
\eta=-b_0 \frac{g^2}{(4\pi)^2} -b_1\, \frac{g^2}{(4\pi)^2}\, w_2,
\label{5.1} 
\end{equation}
with some coefficient $b_1$, and $b_0$ being the correct one-loop
result. Obviously, choosing furthermore the truncation
$w_2,w_3,\dots=0$ leaves us with a purely perturbative lowest-order
result. This means that all nonperturbative information is contained
in the flow of $w_2$, which in turn can be reliably computed only by
including $w_3$, etc. A good estimate therefore probably requires a
very large truncation. Even if the precise infrared values of the
higher couplings $w_i$ may not be very important, their flow exerts a
strong influence on the running coupling in this approximation.

Let us now take the $\pat Z_k$ terms into account, but still neglect
the $\pat\Gt$ terms. In this case, the flow equation results in the
following expression for the anomalous dimension:
\begin{equation}
\eta=-\frac{b_0 \frac{g^2}{(4\pi)^2} +b_1\, \frac{g^2}{(4\pi)^2} w_2}
{1+d_1\,\frac{g^2}{(4\pi)^2}+d_2\,\frac{g^2}{(4\pi)^2} w_2},
\label{5.2}
\end{equation}
with further coefficients $d_1,d_2$, where $d_1<0$. Particularly this
$d_1$ makes an important contribution to the two-loop $\beta$ function
coefficient.  Contrary to Eq.\re{5.1}, this equation contains
information to all orders in $g^2$, even for the strict truncation
$w_2,w_3,\dots=0$. We have to conclude that an adjustment of the
cutoff function using a cutoff wave-function renormalization $Z_k$ is
an effective way to put essential information of the flow of the
higher couplings $w_2,w_3\dots$ into $\eta$. In other words, the
truncated RG trajectory better exploits the degrees of freedom left in
the truncation. Let us note in passing that the flow governed by
Eq.\re{5.2} runs into a kind of Landau pole for
$\frac{g^2}{(4\pi)^2}\simeq 1/|d_1|$, even if the flows of $w_2$ and
higher couplings are included. This ``disease'' has occurred in many
flow equation studies in Yang-Mills theory \cite{Reuter:1993kw},
\cite{Ellwanger:1995qf}, \cite{Reuter:1997gx},
\cite{Bergerhoff:1997cv}.

Now let us turn to the full flow equation, including the terms
generated by $\pat\Gt$. As explained in the previous section, the
right-hand side cannot be displayed in terms of the $w_i$'s in closed
form, because infinitely many terms contribute. Even if we set all
$w_i$'s to zero, which indeed corresponds to our final approximation,
the anomalous dimension reads
\begin{equation}
 \eta=-\frac{b_0 \frac{g^2}{(4\pi)^2} +b_1\, \frac{g^4}{(4\pi)^4}+b_2
   \, \frac{g^6}{(4\pi)^6} +\dots}
{1+d_1\,\frac{g^2}{(4\pi)^2}+d_2\,\frac{g^4}{(4\pi)^4}+d_3
   \, \frac{g^6}{(4\pi)^6} +\dots },
\label{5.3}
\end{equation}
with some real coefficients $b_i$ and $d_i$; (expanding Eq.\re{5.3} in
powers of $g^2$ results in Eq.\re{4.9}). Whereas the nonperturbative
dependence of $\eta$ in Eq.\re{5.2} resembles that of a Dyson series
and is controlled by one coefficient ($d_1$ in that case), Eq.\re{5.3}
contains nonperturbative information from infinitely many
coefficients. The latter arises from the flows $\pat w_i$ which all
contribute to Eq.\re{5.3}. We conclude that the spectrally adjusted
cutoff provides for an efficient reorganization of the flow equation,
so that a small truncation can contain information which, for ordinary
cutoffs, is distributed over infinitely many couplings of a larger
truncation. From this observation, we conjecture that the spectrally
adjusted cutoff selects a truncated RG trajectory which is
``optimized'' with respect to the degrees of freedom within a chosen
truncation. We furthermore conjecture that this trajectory does not
flow into regions of theory space where the exact flow would be mainly
driven by couplings which are not contained in the truncation, but is
always driven by the couplings within the truncation in an optimized
way. Whether the truncated RG trajectory flows to the true quantum
action or not depends, of course, on the quality of the truncation.

We have obviously verified these conjectures only for a truncation
($\eta$) within a truncation ($\eta,w_i$). In fact, in order to
exploit the properties of the spectrally adjusted cutoff, we first
have to discuss the flow of a larger truncation, then disentangle the
flows of the single couplings and finally restrict the calculation to
the most relevant part under consideration.

Let us finally point out that using the spectrally adjusted cutoff
necessarily requires introducing a background field, because $\Gt$ in
the cutoff function is not allowed to depend on the actual field
variable. A background field generally complicates the formulation,
and the technical advantages of the spectrally adjusted cutoff may be
compensated for by these further complications. Gauge theories,
however, may serve as a natural testing ground for the spectrally
adjusted cutoff, since the background-field formalism is advantageous
here for further reasons.

\section{Conclusions}
\label{conclusions}

Starting from the exact renormalization group flow equation for the
effective average action in SU($N$) Yang-Mills theory in $d$
dimensions, we derive within a series of systematic approximations the
$\beta$ function of the gauge coupling. In $d=4$ spacetime dimensions,
the resulting flow of the gauge coupling exhibits accurate
perturbative behavior and approaches a fixed point in the infrared.
The fixed-point results are displayed in Eqs.\re{4.30} and\re{4.32}
for gauge groups SU(2) and SU(3).

In view of the approximations involved, a number of improvements are
desirable in order to confirm the existence of the infrared fixed
point. Above all other possible improvements, such as enlarged
truncations and explicit cutoff-shape independence (or insensitivity),
a better control of gauge invariance under the flow is necessary.

Nevertheless, in view of the flow equation studies performed in the
literature for gauge theories so far, it is already remarkable that
our approximation to the exact flow equation is integrable down to
$k\to 0$; in many instances, the truncation revealed an explicit
insufficiency by developing a Landau-pole type of singularity at some
finite $k$ in one or more couplings. The new technique in the present
work is the use of a cutoff function that adjusts itself permanently
to the actual spectrum under the flow. From a practical viewpoint,
this cutoff condenses information, which is usually distributed over
the flow equations of infinitely many couplings, into the flow
equation of a single coupling (in this case the gauge coupling). We
have reason to believe that the information, which is reorganized in
this way into a single flow equation, is the relevant information that
mainly drives the flow of the corresponding coupling. The fact that we
improved the agreement with the perturbative two-loop running from
merely 113\% to 99\% for SU(2) (using the exponential cutoff shape
function) may serve as a hint in this direction.

If the fixed point exists and our truncation even covers the true
mechanism, it is still unlikely that our present results for
$\alpha_\ast$ are also quantitatively correct. We expect a lowering of
$\alpha_\ast$ for larger truncations owing to the following argument:
in our calculation, the position of $\alpha_\ast$ is strongly governed
by those modes which are also responsible for asymptotic freedom
(contained in $\eta_{\text{b}}$). If, in a larger truncation, operators
of higher order are generated under the flow, these modes will
generically lose influence, and the effects of the remaining spectrum
contributing to $\eta_{\text{a}}$ will be enhanced. This will shift
$\alpha_\ast$ to smaller values. 

A similar effect occurs upon the inclusion of quark degrees of
freedom. The perturbative quark contribution to the $\beta$ function
is already positive. And since no ultraviolet stable fixed point is
known in QED, we also do not expect negative quark contributions
beyond the perturbative regime. Therefore, we expect not only the
presence of the fixed point in full QCD, but also a substantial shift
towards lower values of $\alpha_\ast$. Work in this direction is in
progress.

A comparison of our result with the literature is in order now,
although it is generally difficult, owing to the various
nonperturbative definitions of the gauge coupling; different
definitions may agree perturbatively, but differ beyond perturbation
theory. Our definition is standard in pure continuum gauge theory; moreover,
it is equal to the interaction strength of static quarks with the gauge
field. Nevertheless, it is not immediately clear to us how it can be
related to a definition which is used, for instance, in lattice
gauge theory \cite{Luscher:1993gh}. This may serve as a word of caution.

The notion of an infrared fixed point for the gauge coupling has been
used extensively in recent years, especially in connection with the
phenomenology of power corrections in QCD \cite{Dokshitzer:1998pt}.
Furthermore, such a so-called freezing of the coupling has been
discussed in phenomenological low-energy models \cite{Eichten:1974af},
and deduced from an analysis of the famous $R_{e^+e^-}$ ratio
\cite{Mattingly:ud}. 

There are also various theoretical arguments favoring an infrared
fixed point, e.g., even within a perturbative framework for a finite
number of flavors \cite{Banks:nn}. Furthermore, investigating
analyticity properties in the time-like and space-like (Euclidean)
region, a scheme called analytic perturbation theory has been
proposed, yielding an infrared finite coupling \cite{Shirkov:1997wi};
this program has been successfully applied to hadron and lepton-hadron
phenomenology \cite{Stefanis:2000vd}. Having the above-mentioned
reservations in mind concerning the various different nonperturbative
definitions of the coupling, the question of how they are related to
each other deserves further study.

Moreover, an actual nonperturbative computation of gluon and ghost
propagators has been set up in the framework of truncated
Schwinger-Dyson equations in Landau gauge \cite{vonSmekal:1997is},
revealing an infrared fixed point; these results also receive some
support from lattice calculations \cite{Bonnet:2001uh}. Again, the
relation to our results is not immediately obvious, since the running
coupling as defined in \cite{vonSmekal:1997is} is obtained from the
ghost-gluon vertex; furthermore, a nonperturbative treatment of the
ghost sector turned out to be crucial in that work, but the four-gluon
vertex was neglected.  Nevertheless, there are also similarities: on
very general grounds, it was found in the approximation of
\cite{vonSmekal:1997is} that the fixed point scales with the number of
colors as $\alpha_\ast\sim1/N$.  We observe that the central value of
our SU(3) result and the SU(2) result fulfil exactly this relation,
although this is far from self-evident in our calculation.

Let us finally discuss further implications of our result: comparing
the full $\beta$ function with its perturbative counterpart, we
observe a quantitative agreement up to $\alpha_{\text{s}}\sim 1$. This
does not, of course, justify the use of perturbation theory up to
$\alpha_{\text{s}}\sim 1$ {\em in general}, but may explain why
perturbation theory gives an accurate answer to {\em some} questions,
even at its validity limit.

Concerning the low-energy fixed-point region, one may ask whether our
result provides for some signals of confinement and an expected mass
gap in gauge theories. In the first place, the answer is no, since a
strong coupling does not necessarily imply confinement. It is rather
likely that the strong coupling of the gauge fields is necessary to
give rise to a change of the effective degrees of freedom. These
degrees of freedom (not necessarily included in our truncation) with
probably nontrivial topological properties will then act as
``confiners''. Also the picture of confinement arising in the
framework of Landau-gauge Dyson-Schwinger equations
\cite{vonSmekal:1997is} cannot be contained in our truncation, since
it is based on an infrared enhancement of the ghosts which are treated
rather poorly in the present work. Improvements in this direction are
also subject to future work. As far as a mass gap is concerned, the
infrared fixed point behavior is compatible with such a gap; this is
because a mass gap cuts off all quantum fluctuations of lower
momentum, so that nothing remains to drive the flow. But the mere
existence of an infrared fixed point does not require a mass gap.

An indirect signal of a mass gap may be found in the analysis of the
different spectral contributions; as we have mentioned above, the
perturbative $\beta$ function is mainly determined by the lowest modes
in the spectrum, i.e., the lowest Landau levels in the
covariant-constant field analysis. As is familiar from QED
calculations, the lowest-Landau-level approximation is always
appropriate if the field strength exceeds the mass of the fluctuating
particle. This is certainly the case in the perturbative domain where
the gluon is massless; hence the picture is complete. When we enter
the infrared fixed-point region, the contributions from the remaining
part of the spectrum $\eta_{\text{a}}$ become important. In the
Landau-level picture, this is always the case if a mass of the order
of the lowest Landau level and beyond is present. The value of the
mass then controls the influence of the remaining spectrum. Therefore,
the influence of the complete spectrum at the fixed point may be a
hint for a hidden new mass scale in low-energy Yang-Mills theory.

\section*{Appendices}

\renewcommand{\thesection}{\mbox{\Alph{section}}}
\renewcommand{\theequation}{\mbox{\Alph{section}.\arabic{equation}}}
\setcounter{section}{0}
\setcounter{equation}{0}

\section{Decomposition of $\Gt$}
\label{appDecomp}

Here we briefly describe the method developed in \cite{Reuter:1997gx}
for decomposing $\Gt$ into smaller building blocks suitable for
further diagonalization. The method is based on the observation that
it is sufficient to consider only a covariant constant magnetic
background field in order to project the flow equation onto the
present truncation.

The method consists of identifying those components of the quantum
fluctuations which are appropriately oriented with respect to the
background field; the latter is chosen to be of the type
\begin{equation}
A_\mu^a=n^a\,\Aa_\mu, \quad 
\Fa_{\mu\nu}=\pd_\mu\Aa_\nu-\pd_\nu\Aa_\mu=B\, \epsilon_{\mu\nu}^\bot
=\text{const.}, \label{A1}
\end{equation} 
where $n^a$ is a constant unit vector in color space, $n^2=1$, and
$\Aa_\mu,\Fa_{\mu\nu}$ denote the ``abelian'' gauge field and field
strength. The constant tensor $\epsilon_{\mu\nu}^\bot$ characterizes
the space directions which are affected by the constant magnetic field
upon the Lorentz force, e.g., $\epsilon_{12}^\bot=-\epsilon_{21}^\bot=1$
for $B$ pointing into the 3 direction.

Let us first define two important operators involving the covariant
derivative $(D_\mu[A])^{ab}=\pd_\mu\delta^{ab}-\I \gb A_\mu^c(T^c)^{ab}$
in the adjoint representation:
\begin{eqnarray}
(\Dt)_{\mu\nu}^{ab}&=&(-D^2 \delta_{\mu\nu} +2\I\gb F_{\mu\nu})^{ab}
\nonumber\\
(\Dl)_{\mu\nu}^{ab}&=&-(D\otimes D)_{\mu\nu}^{ab}\equiv -D_\mu^{ac}
D_\nu^{cb}, \label{A2}
\end{eqnarray}
where $(F_{\mu\nu})^{ab}=F_{\mu\nu}^c (T^c)^{ab}$. For covariant
constant fields of the type \re{A1} satisfying the equations of motion
$[D_\mu,F_{\mu\nu}]=0$, the operators $\Dt$ and $\Dl$ commute. As a
consequence, projection operators can be introduced:
\begin{equation}
\Pl=\Dt^{-1}\Dl, \quad \Pt=1-\Pl, \label{A3}
\end{equation}
which obey $P_{\text{T,L}}^2=P_{\text{T,L}}$, $\Pt+\Pl=1$,
$\Pt\Pl=0=\Pl\Pt$. The subscripts indicate that these projectors
reduce to the standard longitudinal and transverse projectors in the
limit $A_\mu\to0$. 

Another pair of projectors can be defined which act solely in color
space:
\begin{equation}
\Pp^{ab}=n^a n^b, \quad \Pb^{ab}=\delta^{ab} -n^a n^b. \label{A4}
\end{equation}
These four projectors are remarkably efficient in the present case;
differentiating our truncation for $\Gk[A,\bA]$, as given in
Eq.\re{3.10}, twice with respect to $A$ and the ghost fields, then
setting $A=\bA$ and dropping the bar, we can represent the result as
\begin{eqnarray}
\Gt[A,A]&=&\Pt\Pb \, \big[W_k'\, \Dt\big] + \Pl \Pb\, \left[
  \frac{1}{\alpha}\, \Dt \right] \nonumber\\
&&+\Pt\Pp\, \big[W_k'\, (-\pd^2)+W_k''\, \mathsf{S}\big]
  +\Pl\Pp\, \left[ \frac{1}{\alpha}\, (-\pd^2)\right]\nonumber\\
&&+P_{\text{gh}}\, \bigl[-D^2], \label{A5}
\end{eqnarray}
where we introduced 
\begin{equation}
\mathsf{S}_{\mu\nu}=\Fa_{\mu\alpha}\Fa_{\beta\nu} \pd^\alpha\pd^\beta
, \label{A5a}
\end{equation}
and $P_{\text{gh}}$ projects trivially onto the ghost sector. 

Equation\re{A5} is perfectly suited for further manipulation, since
the spectra of the operators occurring in the square brackets is
known. This decomposition also offers the possibility of conveniently
implementing different wave-function renormalization constants for
each subcomponent.

\section{Heat-kernel computations}
\label{heatkernel}
\setcounter{equation}{0}

In this appendix, we summarize the results for the heat-kernel traces
appearing in Eq.\re{3.14}. Again, it is sufficient to perform the
calculation for a covariant constant background field in order to
disentangle the contributions to the flow of different operators. 

Let us first mention that all color traces occurring in Eq.\re{3.14}
are of the form
\begin{equation}
\trc\, f\big(n^c\, (T^c)^{ab}\big)=\sum_{l=1}^{N^2-1} f(\nu_l),
\label{B1}
\end{equation}
where $f$ is an arbitrary function, and $\nu_l$ denotes the
eigenvalues of the matrix $(n^c\, T^c)^{ab}$. 

We begin with the heat-kernel trace involving the Laplacian in the
covariant constant magnetic background; the spectrum is given by 
\begin{equation}
\text{Spect.} D^2:\quad q^2+(2n+1)\Bl, \quad \Bl=\gb|\nu_l|B
,\quad n=0,1,\dots,\label{B2}
\end{equation}
where $q_\mu$ denotes the $(d-2)$ dimensional Fourier momentum in
those spacetime directions which are not affected by the magnetic
field. The index $n$ labels the Landau levels; their corresponding
density of states is $\Bl/(2\pi)$. Tracing over the spectrum, we
obtain
\begin{equation}
\frac{1}{\Omega}\, \Tr_{x\text{c}}\, \E^{-\lambda(-D^2)}
=\sum_{l=1}^{N^2-1} \frac{2}{2(4\pi)^{d/2}}\,
\frac{1}{\lambda^{d/2}}\, \frac{\lambda \Bl}{\sinh
  \lambda\Bl}. \label{B3} 
\end{equation}
Here, $\Omega$ denotes the spacetime volume. With reference to
Eq.\re{3.14}, the parameter $\lambda$ can be identified with
$\lambda=sW_k'/(Z_k k^2)$ or $\lambda=s/k^2$.

Next, we turn to the heat-kernel trace involving the operator $\Dt$ as
defined in Eq.\re{A2}. The spectrum is given by
\begin{eqnarray}
&&q^2+(2n+1)\Bl, \quad \text{multiplicity } (d-2)\nonumber\\
\text{Spect.} \Dt:&&q^2+(2n+3)\Bl, \quad \text{multiplicity }
1\label{B4}\\ 
&&q^2+(2n-1)\Bl, \quad \text{multiplicity } 1,\nonumber
\end{eqnarray}
with $q$ and $n$ as in Eq.\re{B2}. The last line contains the
Nielsen-Olesen unstable mode for $n=0$ \cite{Nielsen:1978rm}, which
has a tachyonic part for small momenta $q^2$. Tracing over the
spectrum, we find
\begin{equation}
\frac{1}{\Omega}\, \Tr_{x\text{cL}}\, \E^{-\lambda\Dt}
=\sum_{l=1}^{N^2-1} \frac{2}{2(4\pi)^{d/2}}\, \frac{1}{\lambda^{d/2}}
\left(d\,\frac{\lambda \Bl}{\sinh \lambda\Bl} 
  +4\lambda\Bl \sinh \lambda\Bl\right).   \label{B5}
\end{equation}
Finally, we need the following traces 
\begin{eqnarray}
\frac{1}{\Omega}\, \Tr_{x}\,
\E^{-\lambda(-\pd^2)}&=&\frac{2}{2(4\pi)^{d/2}}
\, \frac{1}{\lambda^{d/2}}, \nonumber\\
\frac{1}{\Omega}\, \Tr_{x\text{cL}}\,
  \E^{-\lambda(-\pd^2)-\lambda'\mathsf{S}}
  &=&\frac{2(d-1)}{2(4\pi)^{d/2}}\, \frac{1}{\lambda^{d/2}}   
  +\frac{2}{2(4\pi)^{d/2}}\, \frac{1}{\lambda^{d/2}} \,
    \frac{\lambda}{\lambda+B^2\,\lambda'}, \label{B6}
\end{eqnarray}
where $\mathsf{S}$ has been defined in Eq.\re{A5a}. Here and in
Eq.\re{B5}, the $\lambda$ parameters abbreviate $\lambda=sW_k'/(Z_k
k^2)$ and $\lambda'=sW_k''/(Z_k k^2)$. 

Equations\re{B3},\re{B5},\re{B6} serve as the main input for
evaluating the right-hand side of the flow equation in
Sect.~\ref{floweq}.

\section{Expansions}
\label{series}
\setcounter{equation}{0}
Here we shall explicitly display the expansions which are required for
the analysis of the anomalous dimension in Sect.~\ref{expansion}. The
series given below are expanded in terms of the renormalized
dimensionless field strength squared $\te$, but they are also related
to expansions in terms of the propertime parameter $s$ or the
renormalized coupling $g^2$. Since we are expanding an integrand and
then interchange integration with expansion, the resulting series can
(and will) be asymptotic, involving strongly increasing coefficients. 

Neglecting all $w_i$'s in the expansion of
$w_k(\te)=\te+w_2\case{\te^2}{2}+w_3\case{\te^3}{6}\dots$, we obtain
for the expansions of the auxiliary functions $f_{1,2,3}$ as defined
in Eq.\re{3.16} (recall that $b_l=|\nu_l|\sqrt{2\te}$):
\begin{eqnarray}
2\!\sum_{l=1}^{N^2-1}\! f_1(s\wdot b_l)\, b_l^{d/2}\bigg|_{w_i\to 0}
\!\!\!\!\!\!\!\!\!
&=&\!\! -(d\!-\!1)\! \sum_{i=0}^\infty \frac{2^i (2^{2i}\!-2)}{(2i)!}
\, \tau_i\,  B_{2i}\, s^{2i-d/2}\, \te^i%, \nonumber\\
%&&
+4\sum_{i=0}^\infty \frac{2^i }{(2i-\!1)!} \, \tau_i\, s^{2i-d/2}\,
  \te^i \nonumber\\
2\sum_{l=1}^{N^2-1} f_2(s b_l)\, b_l^{d/2} 
&=& - \sum_{i=0}^\infty \frac{2^i (2^{2i}-2)}{(2i)!} \, \tau_i\,
  B_{2i}\, s^{2i-d/2}\, \te^i, \label{E2}
\end{eqnarray}
where $B_{2i}$ denotes the Bernoulli numbers, and we define
$1/(-1)!=0$. The $\tau_i$ are defined in Appendix \ref{SU3} and are
related to the group theoretical factors $\sum_{l=1}^{N^2-1}
(\nu^2)^i$ that occur in the expansions given above. Whereas the
expansion of $f_3$ vanishes in the present approximation, the
expansion of its derivatives, as they occur in the last line of
Eq.\re{3.20}, must be retained:
\begin{equation}
(\pat-4\te\pd_\te+d)\,
f_3\left(s\wdot,\frac{\wdot}{\wdot+2\te\wddot}\right)\bigg|_{w_i\to0}
=\sum_{i=1}^{\infty} \frac{2i}{s^{d/2}}\, \frac{\te^i}{i!}\, \pat
w_{i+1}. \label{E3}
\end{equation}

\section{Cutoff functions}
\label{cutoff}
\setcounter{equation}{0}

In Eq.\re{3.2}, we introduce the cutoff function
$R_k(x)=x\,r(\case{x}{Z_k k^2})$, where $r(y)$ is a dimensionless
function of a dimensionless argument. For actual computations, we
need the combinations $h(y)$ and $g(y)$ as well as their Laplace
transforms $\hi$ and $\gt$ as defined in Eqs.\re{hg} and\re{LThg}. 

Instead of choosing a certain cutoff function by specifying $r(y)$, we
can specify a function $h(y)$, or alternatively $g(y)$, which fixes the
remaining functions by virtue of Eq.\re{hg}; the direct connection
between $h(y)$ and $g(y)$ can be formulated as
\begin{equation}
y\frac{d}{dy} g(y)=\bigl(g(y)-1\bigr)\,  h(y). \label{D1}
\end{equation}
A similar reasoning holds for a definition of the cutoff in Laplace
space by specifying one of the functions $\hi$ or $\gt$, for which
Eq.\re{D1} translates into
\begin{equation}
\gt+s\frac{d}{ds}\gt=\hi-\int_0^s dt\,
\tilde{h}(t)\,\tilde{g}(s-t). \label{D2} 
\end{equation}
These identities can be used to define a desired cutoff in its
simplest representation without the need to specify the corresponding
function $r(y)$ explicitly; the latter might look very complicated. Of
course, one has to take care of all the necessary conditions that a
cutoff has to satisfy as listed in Eqs.\re{3.3} and\re{3.4}.

During the expansion of the propertime integrand in
Sect.~\ref{expansion}, we encounter the moments of $\hi$ and $\gt$ as
defined in Eq.\re{4.10}. These moments can also be translated into
a momentum space calculation (``$y$ space''):
\begin{eqnarray}
h_{-j}:=\int\limits_0^\infty \frac{ds}{s^j}\, \hi
&=&\frac{1}{\Gamma(j)}\int\limits_0^\infty dy\, y^{j-1}\, h(y), \quad
  j>0,\nonumber\\
h_j:=\int\limits_0^\infty {ds}\,s^j\, \hi
&=&\lim_{y\to 0} (-1)^j\, \left(\frac{d}{dy}\right)^{(j)}
h(y),\quad j\geq0 \label{D3}
\end{eqnarray}
and equivalently for the $g_j$'s.

In this work, the exponential cutoff is technically advantageous; all
functions involved have a simple representation:
\begin{eqnarray}
r(y)&=&\frac{1}{\E^y -1}, \quad h(y)=\frac{y}{\E^y-1}, \quad g(y)=\E^{-y},
\nonumber\\
\hi&=&-\sum_{m=1}^\infty \delta(s-m)\, \frac{d}{ds}, \quad 
  \gt=\delta(s-1), \label{D4}
\end{eqnarray}
where the $s$ derivative acts on the remaining propertime
integrand. For the moments required in $d=4$, we find
\begin{eqnarray}
g_j&=&1,\nonumber\\
h_{-2}&=&2\, \zeta(3)\simeq 2.404\dots,\label{D5}\\
h_j&=&B_{j},\quad j=1,2,\dots,\nonumber
\end{eqnarray}
where $B_{j}$ symbolizes the Bernoulli numbers.

\section{SU(2) versus SU(3)}
\label{SU3}
\setcounter{equation}{0}

Gauge group information enters the flow equation via the color traces.
In Appendix~\ref{heatkernel}, we evaluated these traces formally by
introducing the eigenvalue of $(n^c\,T^c)^{ab}\to \nu_l$,
$l=1,\dots,N^2-1$. During the expansion of the right-hand side of the
flow equation in Sect.~\ref{expansion}, we encounter the following
factors:
\begin{equation}
\sum_{l=1}^{N^2-1} \nu_l^{2i}=n^{a_1} n^{a_2} \dots n^{a_{2i}}
\, \trc [T^{(a_1} T^{a_2}\dots T^{a_{2i})}],\label{C1}
\end{equation}
where the parentheses at the color indices denote symmetrization. For
general gauge groups, these factors are not independent of the
direction of $n^a$. Contrary to this, the left-hand side of the flow
equation is a function of $\case{1}{4} F_{\mu\nu}^a F_{\mu\nu}^a\to
\case{1}{2} B^2$, which is independent of $n^a$. Therefore, we do not
need the complete factor of Eq.\re{C1}, but only that part of the
symmetric invariant tensor $\trc [T^{(a_1}\dots T^{a_{2i})}]$ which is
proportional to the trivial one:
\begin{equation}
\trc [T^{(a_1} T^{a_2}\dots T^{a_{2i})}]=\tau_i\, \delta_{(a_1
  a_2}\dots\delta_{a_{2i-1} a_{2i})}+\dots, \label{C2}
\end{equation}
where we omitted further nontrivial symmetric invariant tensors. These
omitted terms do not contribute to the flow of $W_k(\theta)$, but to
the flow of other operators which do not belong to our truncation,
e.g., operators involving contractions of the field strength tensor
with the $d_{abc}$ symbols. 

For SU($N$) gauge groups, we trivially deduce that
\begin{equation}
\tau_0=N^2-1,\quad \tau_1=N. \label{C3}
\end{equation}
For the gauge group SU(2), all complications are absent, since there
are no further symmetric invariant tensors in Eq.\re{C2}, implying
\begin{equation}
\tau^{N=2}_i=2, \quad i=1,2,\dots\,\,.\label{C4}
\end{equation}
For the gauge group SU(3), we do not evaluate the $\tau_i$'s from
Eq.\re{C2} directly; instead, we exploit the fact that the color unit
vector can always be rotated into the Cartan subalgebra. For SU(3), we
choose a color vector $n^a$ pointing into the 3 or 8 direction in
color space, representing the two possible extremal cases:
\begin{equation}
\tau^{N=3}_{i,3}=2+\frac{1}{2^{2i-2}},\quad 
\tau^{N=3}_{i,8}=\frac{3^i}{2^{2i-2}}. \label{C5}
\end{equation}
Note that their limiting behavior is rather different: for
$i\to\infty$, we find $\tau^{N=3}_{i,3}\to 2$, but
$\tau^{N=3}_{i,8}\to 0$.

The uncertainty introduced by the artificial $n^a$ dependence of the
color traces is finally responsible for the uncertainty of our result
for the SU(3) infrared fixed point.

\section{Numerical computations}
\label{numerics}
\setcounter{equation}{0}

Since the numerical evaluation of the anomalous dimension $\eta$
depending on the coupling $G=g^2/[2(4\pi)^2]$ as represented in
Eqs.\re{4.23} and\re{4.25} is not straightforward, we mention here some
details about the multidimensional integration and summation. We begin
with the part $\eta_{\text{a}}$ in Eq.\re{4.23}: substituting
$s_1/s_2\to s_1$, the $s_2$ integral can be performed, resulting in
the modified Bessel function $K_{3N^2-4}(2\sqrt{s_1})$. Substituting
furthermore $t\to t/l$, and defining the expressions
\begin{equation}
L(t):=\sum_{l=1}^\infty \frac{1}{2} \frac{1}{1+\cosh lt}\, \frac{1}{l},
\quad
\widetilde{K}(s_1):=s_1^{3N^2/2-1}\, K_{3N^2-4}(2\sqrt{s_1}),
\label{F1}
\end{equation}
we obtain the representation
\begin{equation}
\eta_{\text{a}}^{N=2}=\frac{64NG}{\Gamma(3N^2\!-2)\pi^2}
\int\limits_0^\infty dt\, L(t)\int\limits_0^\infty ds_1\,
\widetilde{K}(s_1)\, \left[ S\left(\frac{cGs_1t^2}{2\pi^4} \right)
-\frac{1}{2}\, S\left(\frac{cGs_1t^2}{8\pi^4}
\right)\!\right]. \label{F2}
\end{equation}
Apart from an easily integrable $1/\sqrt{t}$ singularity induced by
$L(t)$, the integrals are smooth and drop off exponentially for large
$t$ and $s_1$ in the required $G$ range. The sum $S(q)$ defined in
Eq.\re{4.24} converges quickly and an accuracy with error $<1\%$
requires only ${\cal O}(100)$ terms or less. The sum $L(t)$ is rather
slowly converging for small $t$, but the same accuracy can be obtained
by including ${\cal O}(10^5-10^6)$ terms. Depending on the actual
value of the arguments $t$ and $q$, we adjust the included number of
terms dynamically.

For the part $\eta_{\text{b}}$, different complications
occur. Beginning with Eq.\re{4.25}, we substitute $s\to
st\sqrt{cG/(2\pi^2)}$ (and similarly for $s'$) and find
\begin{equation}
\eta_{\text{b}}^{N=2}=-\frac{32NG}{\Gamma(3N^2\!-2)}\, \text{Re}
\int\limits_0^\infty \case{(1\!+\I)}{\sqrt{2}}dt 
  \frac{\E^{\frac{1+\I}{\sqrt{2}} t}}{(\E^{\frac{1+\I}{\sqrt{2}}
  t}+1)^2}\, I_s \left( \sqrt{ \frac{cG}{2\pi^2}}\, t\right),
\label{F3}
\end{equation}
where we defined
\begin{equation}
I_s \left( x\right) =
\frac{1}{x}
\int\limits_0^{x} ds \left(\frac{d}{ds}\right)^{(3N^2-3)}
\frac{d}{ds'} \, s^{3N^2-3} s'\, S(-\I ss')\Big|_{s'=x-s}. \label{F4}
\end{equation}
The problem here is that the derivatives cannot be carried out
numerically with a sufficient accuracy, but have to be computed
analytically within the sum representation for $S(-\I ss')$. This
implies that each term in the sum then consists of $\sim 20$ terms
for SU(2) and $\sim 50$ for SU(3). This limits the generalization of
the calculation to higher gauge groups for technical reasons. The
remaining $s$ and $t$ integrations can easily be performed to a high
accuracy. We estimate the total error of the numerical computation to
be within a few percent.

\section*{Acknowledgment}
The author would like to thank D.F.~Litim and J.M.~Pawlowski for
numerous discussions, for comments on the manuscript, and for
communicating their results of Refs.~\cite{Litim:2001hk}
and~\cite{Litim:2002} prior to publication. The author is also
grateful to R.~Alkofer, W.~Dittrich, G.V.~Dunne, C.S.~Fischer,
K.~Langfeld, J.I.~Latorre, S.~Sint and C.~Wetterich for helpful
information and correspondence, and he wishes to thank W.~Dittrich for
carefully reading the manuscript. This work is supported by the
Deutsche Forschungsgemeinschaft under contract {Gi 328/1-1}.

\end{document}